\documentclass[pdflatex,sn-mathphys-num,twocolumn]{sn-jnl}

\usepackage{graphicx}%
\usepackage{multirow}%
\usepackage{amsmath,amssymb,amsfonts}%
\usepackage{amsthm}%
\usepackage{mathrsfs}%
\usepackage[title]{appendix}%
\usepackage{xcolor}%
\usepackage{textcomp}%
\usepackage{manyfoot}%
\usepackage{booktabs}%
\usepackage{algorithm}%
\usepackage{algorithmicx}%
\usepackage{algpseudocode}%
\usepackage{listings}%
\usepackage{siunitx}
\usepackage{float}

\theoremstyle{thmstyleone}%
\theoremstyle{thmstyletwo}%
\theoremstyle{thmstylethree}%
\begin{document}
\title[Article Title]{Influence of Flow on Discharge Behaviors\\
and CO$_2$ Conversion in Gliding Arc Discharges}


\author[1]{\fnm{} \sur{Alexander Davis}}\email{dralexander@u.northwestern.edu}

\author[2]{\fnm{} \sur{Charles Burton}}\email{charlesburton2028@u.northwestern.edu}

\author[1]{\fnm{} \sur{Stephanie Pecaut}}\email{stephaniepecaut2028@u.northwestern.edu}

\author[3]{\fnm{} \sur{Matthew Hershey}}\email{matthewhershey2026@u.northwestern.edu}

\author[2]{\fnm{} \sur{Michelle M. Driscoll}}\email{michelle.driscoll@northwestern.edu}

\author[1]{\fnm{} \sur{Linsey C. Seitz}}\email{linsey.seitz@northwestern.edu}

\author*[1,3]{\fnm{} \sur{Dayne F. Swearer}}\email{dayne@northwestern.edu}

\affil*[1]{\orgdiv{Department of Chemical Engineering}, \orgname{Northwestern University}} 

\affil[2]{\orgdiv{Department of Physics and Astronomy}, \orgname{Northwestern University}}

\affil*[3]{\orgdiv{Department of Chemistry}, \orgname{Northwestern University}}

\abstract{Plasma reactors present themselves as a unique means of electrified chemical production, particularly in the conversion of greenhouse gases (e.g., CO$_{2}$) back into chemical feedstocks. Warm plasmas, such as gliding arc discharges, represent a growing class of catalyst-free plasma reactors that demonstrate high energy efficiency and scalability. Here, we introduce mean discharge time as a characteristic descriptor of gliding arc dynamics, extracted directly from voltage and current waveforms. Electrical signatures for distinct plasma behaviors (modes) were established using high-speed photography, and discharge time distributions were evaluated as a function of Reynolds number. Mean discharge time is shown to decrease non-linearly with Reynolds number, providing a link between arc behavior and underlying fluid dynamics, which is otherwise neglected in traditional reactor characterization. Mean discharge time thereby provides a quantitative, operando descriptor of transient gliding arc dynamics and stability, offering insight into discharge behavior that traditional characterization approaches do not capture.

}


\maketitle

\section{Introduction}\label{sec1}
Chemical reactors that employ renewable electricity offer a promising pathway for reducing global emissions and enabling modular and rapidly scalable operation in response to demand or power availability.~\cite{bistline_role_2021, bouckaert_net_2021,deangelo_energy_2021, mallapragada_decarbonization_2023,cresko_us_2022} As a subset of electrified technologies, plasma reactors stand out for their scalability, short steady-state times, and reactor diversity. ~\cite{osorio-tejada_co2_2024,george_review_2021,snoeckx_plasma_2017} A plasma is a partially or fully ionized gas, generated for example by applying a potential difference between two electrodes. Plasmas consist of a reactive mixture of free electrons, ions, excited molecular states, radicals, and, depending on the degree of ionization, neutral chemical species. ~\cite{fridman_plasma_2008} Plasmas are broadly classified as equilibrium (thermal) or nonequilibrium (non-thermal) based on the relative temperatures of the bulk gas and electrons, with an intermediate "warm" plasma classification having been recently distinguished.\cite{adamovich_2022_2022}  In each regime the electron temperatures sit around 10$^4$ K, while the gas temperature increases from near ambient in the non-thermal case, to 10$^3$  K, and 10$^4$ K in warm and thermal regimes respectively, as the plasmas approach local thermal equilibrium.\cite{omodhrain_upscaling_2024}

\begin{figure*}[t]
\centering
\includegraphics[width=0.9\linewidth,keepaspectratio]{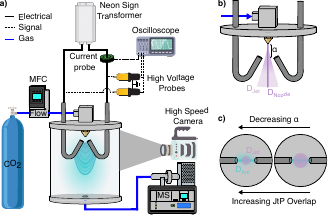}
\caption{
(a) Schematic of the GAD reactor and diagnostic equipment. (b) Jet expansion through the conflat nozzle into the reactor (D$_{Jet}$). (c) Cross-sectional view of the reactor illustrating JtP overlap.
}
\end{figure*}

Plasmas span a wide range of gas temperatures, of which gliding arc discharges (GADs) represent one prominent class of warm plasmas. GADs themselves encompass a broad collection of geometries, with small arc volumes producing intensely thermal cores with steep spatial gradients during arc propagation, and larger arc volumes yielding cooler gas temperatures with shallower thermal gradients, where reactivity is driven by electron-collision and vibrational pathways rather than thermal.~\cite{wang_gliding_2017,czernichowski_spectral_1996} 

These thermal and non-thermal regions of GADs can drive challenging chemical reactions without a catalyst. This unique behavior of GADs has been leveraged for an ever-growing array of useful chemical applications, such as NO$_x$ formation from air, volatile organic compound decomposition, wastewater treatment, and waste gas-to-feedstock reforming. \cite{wang_nitrogen_2017,van_raak_numbering_nodate,gong_decomposition_2020,hameedl_gliding_2020} Among these, CO$_2$ conversion has emerged as a particularly compelling target as the strong C=O bond presents activation challenges that neither purely thermal nor non-thermal plasmas can address efficiently, motivating the utilization of a warm plasma to combine the attributes of both.

Performance in gliding arc reactor systems relies heavily on cyclic arc formation, propagation, and extinction, thus making the control of arc dynamics paramount to GAD operation. Despite extensive research into GAD reactors, the field has predominantly characterized reactor performance through production-related metrics: conversion, production rate, and energy efficiency. This approach largely overlooks the underlying plasma discharge behavior during operation, which is directly connected to such product-related metrics.

Voltage drops in the electrical waveform have been proposed to correspond to arc formation and extinction events, but existing studies lack sufficient temporal resolution to fully define these signals.\cite{potocnakova_experimental_2017,mutaf-yardimci_thermal_2000} As a result, arc dynamics are largely described qualitatively in relation to flow, and few established quantitative descriptors have been introduced to compare across geometries, flow conditions, power supplies, and independent scientific studies. 

Here, we leverage CO$_2$ disproportionation in a modular gliding arc discharge as a model system to gain deeper insights into how plasma discharge behavior correlates with production metrics. We demonstrate a method to extract arc-formation events directly from electrical waveforms and construct distributions of arc-formation times, thereby providing quantitative descriptors of gliding arc dynamics that can be used to scale metrics for future designs. Using time-gated high-speed photography of the CO$_2$ disproportionation reaction, we define electrical signatures of arc initiation, propagation, and extinction events. From these signatures, arc initiations can be tracked to identify distributions of arc formation times. Critically, this approach enables operando characterization of the frequency and variability of arc formation without requiring high-speed imaging beyond the initial signature development. We use this framework to investigate how reactor geometry and gas flow govern arc dynamics, providing insight into the relationship between arc behavior and underlying fluid dynamics for an emerging and promising electrified reactor technology.\cite{bryssinck_performance_2025,omodhrain_upscaling_2024,luo_stabilizing_2025}

\section{Results}\label{sec2}
\subsection{Flow-Dependent Reactor Performance}

A schematic of the experimental setup is shown in \textbf{Figure 1a}. Two diverging aluminum electrodes form a minimum gap of 3.2 mm and open outward at 25° with a bent arm length of 8 mm. Pure CO$_2$ is injected through a nozzle upstream of this gap to produce a jet that drives arcs along the path of the electrodes until the existing arc destabilizes and a new one forms, creating a continuous cycle. 

Three parameters were varied independently in order to control the flow environment experienced by the plasma arc: flow rate, nozzle diameter (D$_{Nozzle}$), and nozzle-to-electrode distance ($\alpha$) as shown in \textbf{Figure 1b}. CO$_2$ conversion was measured via mass spectrometry and plasma power was obtained from voltage and current waveforms as described in \textbf{S1.1}. A detailed assessment of all equipment is provided in the Methods section.

From the reactor geometry, the fraction of the inlet jet envelope (D$_{Jet}$) that intersects with an arc formed at the minimum electrode gap can be calculated from the D$_{Nozzle}$, distance $\alpha$, and fundamental momentum transport equations outlined in \textbf{S1.2}.\cite{pope_turbulent_2000} This geometric overlap, denoted Jet-to-Plasma (JtP) overlap and shown in \textbf{Figure 1c}, provides an estimated upper limit on how much of the inlet gas could be exposed to the plasma. 

Conventional engineering heuristics suggest that exposing more of the gas to the plasma environment will result in higher conversion. The data presented in \textbf{Figure 2} shows no such positive relationship; instead, across both flow rates, increasing JtP overlap fails to uniformly translate into increased conversion, either decreasing or remaining fixed depending on conditions tested. This same trend has previously been reported by Li et al. for a similar system.\cite{li_plasma-assisted_2019} These observations indicate that increasing the exposure between the inlet gas and the assumed arc volume does not intrinsically predict conversion, motivating the analysis of arc behavior in conjunction with flow and reactor geometry. 

We find that our non-optimized reactor design leads to single-pass conversions of $\approx$ 5 \% and energy efficiencies over 20 \%  without the need for catalyst integration. This low-power operation, under 60 W, results in energy costs directly competitive with existing literature, demonstrating that this model system maintains performance characteristics comparable to pioneering GAD reactor designs; tabulated energy costs and comparisons to literature are provided in \textbf{S1.1}.  

\begin{figure}[h]
\centering
\includegraphics[width=0.9\linewidth,keepaspectratio]{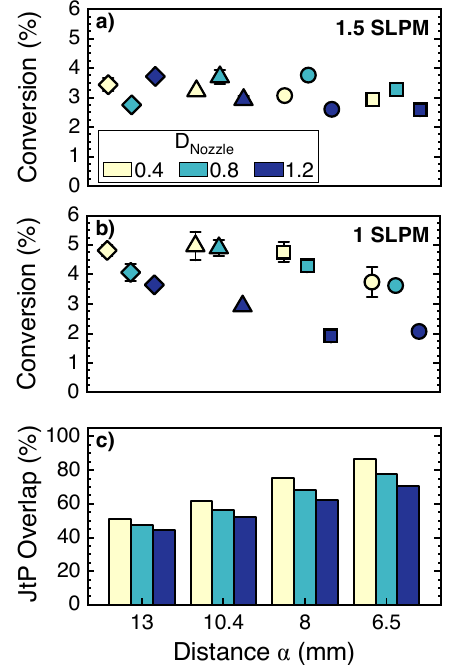}
\caption{
Impact of reactor geometry and flow rate on CO$_2$ conversion, where increased jet-to-plasma (JtP) overlap does not consistently improve conversion. (a–b) CO$_2$ conversion (\%) at flow rates of 1 and 1.5 SLPM, respectively. (c) Percentage of entering CO$_2$ intersecting with a plasma arc formed at the minimum electrode gap for different nozzle diameters (0.4, 0.8, 1.2 mm) and electrode-to-nozzle distances. 
}
\end{figure}

While flow rate is essential for determining production rate and energy efficiency metrics for GAD reactors, its physical relation to a driving force behind plasma arc propagation depends on the geometry of the underlying reactor system. The Reynolds number (Re), the ratio of inertial to viscous forces in the flow, provides a more general descriptor for these systems by combining gas properties, operating conditions, and reactor geometry into a single value for this analysis. However, because gas density and viscosity are temperature-dependent parameters, the choice of reference location and, equivalently, temperature is essential to calculate the Re, due to the large thermal gradients present in GADs. 

\begin{figure}[t]
\centering
\includegraphics[width=0.9\linewidth,keepaspectratio]{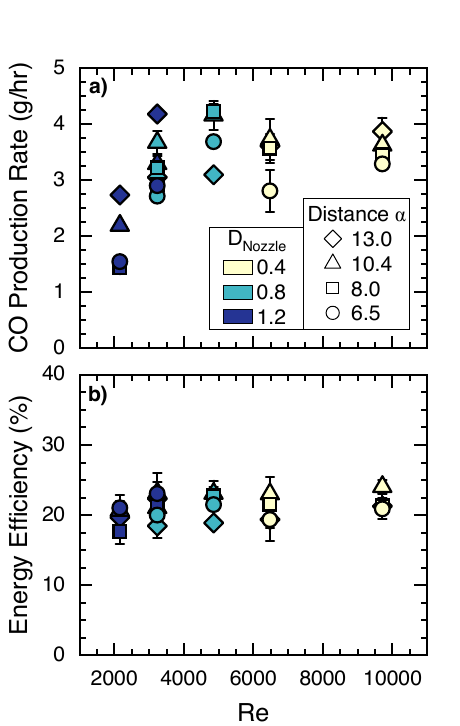}
\caption{
(a) Production rate variance decreases at high Re. (b) Energy efficiency remains largely constant across all tested conditions. Error bars smaller than marker size hidden for clarity.
}
\end{figure}

This work defines Re immediately after the inlet nozzle, where there is no appreciable heating of the incoming gas because it is located upstream of the plasma, providing a metric independent of the arc's temperature and location and enabling direct comparison across all data. From the various nozzles tested, a wide range of Re can be achieved by varying nozzle diameters rather than flow rate alone, allowing this study to span flow conditions similar to those of larger-scale studies in the literature.

\textbf{Figure 3} plots reactor performance against Re,  with \textbf{Figure 3a} showing the sensitivity of production rate on electrode-to-nozzle distance, where an inter-Re variation of up to 1.5 g/hr is observed for most of the data. However, at the largest Re this variation reduces to below 0.6 g/hr with the data clustering toward 3.5 g/hr of CO production.  Additionally, the energy efficiency in \textbf{Figure 3b} remains effectively constant across the full range of Re tested. This is particularly surprising given that the spread of production rates varies by a factor of 1.5 in some cases. 

This result reflects substantially different arc dynamics, resulting in a wide spread of power consumption values to maintain a fixed energy efficiency across the varying production rates. This suggests that different reactor geometries, even at the same Re, can lead to different arc dynamics while preserving energy efficiency, motivating a deeper characterization of the plasma itself.

Jointly, these results show that fluid dynamics play a governing role in the GAD performance, but collective metrics such as conversion, production rate, and energy efficiency are insufficient to directly describe the arc dynamics, motivating the development of a quantitative method to describe these behaviors.

\subsection{High-Speed Identification of Discrete Discharge Modes}

\begin{figure*}[h]
\centering
\includegraphics[width=0.9\linewidth,keepaspectratio]{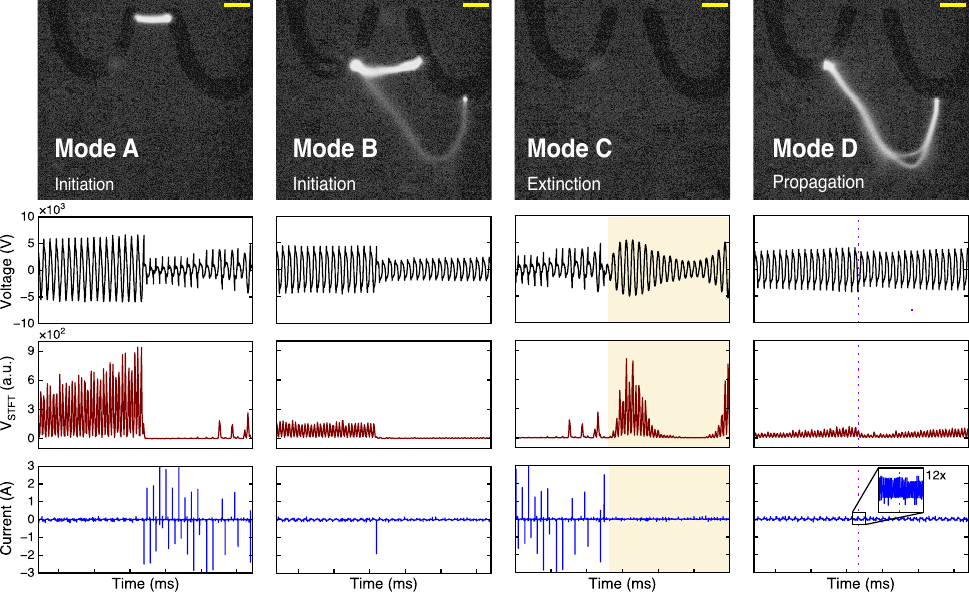}
\caption{ 
Distinct gliding arc discharge modes and their electrical signatures in CO$_2$, with identical y-axis scales across signal types. Major tick marks on the x-axis represent 250 $\mu$s. High-speed photography reveals visually and electrically distinguishable discharge behaviors: initiation events are identified by coincident voltage drops and current fluctuations, while propagation and extinction are characterized by voltage drops alone. Yellow scale bars represent 3.2 mm. }
\end{figure*}

To begin classifying arc dynamics, prior high-speed imaging studies of GADs have identified two general modes, denoted Mode A and Mode B. The modes are distinguished by whether arc formation occurs at the minimum electrode gap or elsewhere.\cite{zhang_mode_2018,zhang_mode_2020, wang_gliding_2017, chen_spatiotemporally_2021, ma_insight_2025,wang_hydroxyl_2022} 

These studies have typically operated at frame rates below 50,000 frames per second (fps), which, for many AC studies, has been slower than the oscillation period of the power supply. This prevents direct time-gating of the imaged arc events to the corresponding waveform in a single measurement, and instead requires reconstructing the arc at each phase of the oscillation from frames captured across many separate cycles. \cite{choi_characterization_2025, zhu_dynamics_2014} 

In this work, high-speed imaging was performed at 200,000 fps, providing a frame interval of \SI{5}{\micro\second}, with an exposure time of \SI{4.53}{\micro\second}, well below the \SI{44}{\micro\second} period of the neon sign transformer (NST) used as a power supply. Importantly, this modern NST was produced in compliance with NEC and UL943 standards requiring the presence of a ground-fault circuit interrupter in the internal circuitry which plays a role in developing these electrical signatures. This temporal resolution enables direct characterization of individual arc events and their corresponding electrical waveforms and, for the first time, reveals two additional discharge modes (designated Modes C and D here) with unique electrical signatures.

\textbf{Figure 4} shows representative high-speed images of discrete modes associated with arc formation, propagation, and extinction events, along with the associated voltage and current waveforms. The V$_{STFT}$ curve is derived from a short-time Fourier transform (STFT) of the raw voltage signal, which analyzes the signal across an overlapping sliding time window rather than an entire waveform at once to amplify local deviations.  This windowed approach yields sharp, characteristic peaks even from subtle voltage deviations while producing clean low-intensity baselines between events; a full derivation is provided in \textbf{S1.6}.

In Mode A, the arc forms at the minimum electrode gap, which, before initiation, is composed of purely neutral gas molecules that must be partially ionized to create the arc. This ionization over a small arc volume produces a high-electron-density arc, thereby permitting a large amount of current to pass until the arc propagates outward and the arc volume grows. \cite{wang_gliding_2017,bourlet_numerical_2022}   The current waveform associated with Mode A shows multiple intense current spikes peaking on the order of several amperes, the duration of which depends on the arc propagation speed as discussed in later sections.

In contrast, Mode B occurs when an arc forms outside the narrowest position of the electrodes and reuses a portion of the ionized path to create a new, more conductive arc further along the discharge gap. As the plasma arc is comprised of excited species, the segments near the electrodes, called the roots, are typically re-utilized to form a new arc (brighter) once the former arc extinguishes (dimmer) in Mode B as these segments move more slowly than other portions of the arc as demonstrated later in \textbf{Figure 5}.  

At the roots, slower-moving gas from the residual ionized and excited species lowers the ionization energy required to establish a new pathway.\cite{zhang_mode_2018,yang_numerical_2023} Because the arc formation in the Mode B case utilizes pre-ionized species from the former arc, and arc volumes remain larger than in the Mode A case, there are fewer instances in which large amounts of current can be passed. The current spike(s) associated with Mode B discharge events are on the order of several amperes down to milliamperes as seen in \textbf{Figure 4}.  These signatures match the Mode A and Mode B descriptions established in prior work, in which formation events are identified by a voltage drop and a change in the current signal.~\cite{zhang_mode_2018,zhang_mode_2020} 

However, the presence of a voltage drop alone is insufficient to identify that a new arc has formed: as shown in \textbf{Figure 4}, both Modes C and D demonstrate voltage drops but have different current signatures relative to Modes A and B, making the simultaneous analysis of voltage and current necessary to distinguish these events from one another.

In contrast to Modes A and B, which represent specific instances in time when a new arc is forming, Mode C represents a range of time, typically about 1 ms, in which no plasma is present. This event is periodic, occurring every $\approx$ 8.33 ms, which is most consistent with the half-period of the 60 Hz wall frequency before conversion to high-voltage, where the presence of the plasma arc triggers a secondary-circuit ground fault protection (SGFP) trip cutting power to the electrical circuit to protect the power supply. This periodic shutoff is an expected consequence of SGFP protections in modern high-voltage equipment and plays a role in controlling arc dynamics.

In this Mode C window, the NST briefly shuts off until the next half-period begins, typically around 1 ms, to reset the fault indicator, and the voltage waveform takes on a distinct shape during the shutdown compared to other modes. The corresponding electrical signature consists of little to no current and several sharp V$_{STFT}$ peaks as the existing arc destabilizes. Importantly, Mode C represents an absence of plasma rather than a discharge event, and its periodic nature serves to provide a maximum lifetime for arcs generated in this GAD system using an accessible NST transformer as a plasma power supply. This difference alone may account for why this mode has not been previously reported in the literature, and has implications for future development of plasma power source design.

The fourth mode identified in this study, Mode D, appears visually similar to Mode B in the high-speed imaging, wherein the morphology of the arc changes, but the new arc retains a very similar shape to the preceding arc. The associated voltage drop is small, often barely distinguishable from the noise level of the raw signal, while the current shows no statistically significant deviation, unlike the milliampere-level deviations possible in Mode B. 

This absence of a current fluctuation is meaningful for understanding the effects of current on gas heating and ionization. The current spikes associated with Modes A and B indicate increased electron density in the newly formed arc, providing the energy to drive chemical reactions forward via both thermal and non-thermal pathways.\cite{wang_gliding_2017,yang_numerical_2023} However, Mode D involves no such energy injection and remains in a low-current, resistive phase throughout the arc restructuring, consistent with the description of arc propagation in the literature. \cite{zhang_mode_2018,zhang_mode_2020}

Following this convention that low current characterizes propagation, Mode D is best treated as a type of propagation behavior rather than a new arc formation, since no variations in current and thus electron density are deposited in the arc to drive additional chemistry. As the voltage drop associated with Mode D is near the noise floor, identifying these events from electrical signals alone is challenging and requires joint optical and electrical diagnostics. Despite this detection challenge, Mode D events are observed across all operating conditions tested.

\begin{figure*}[!t]
\centering
\includegraphics[width=0.9\linewidth, keepaspectratio]{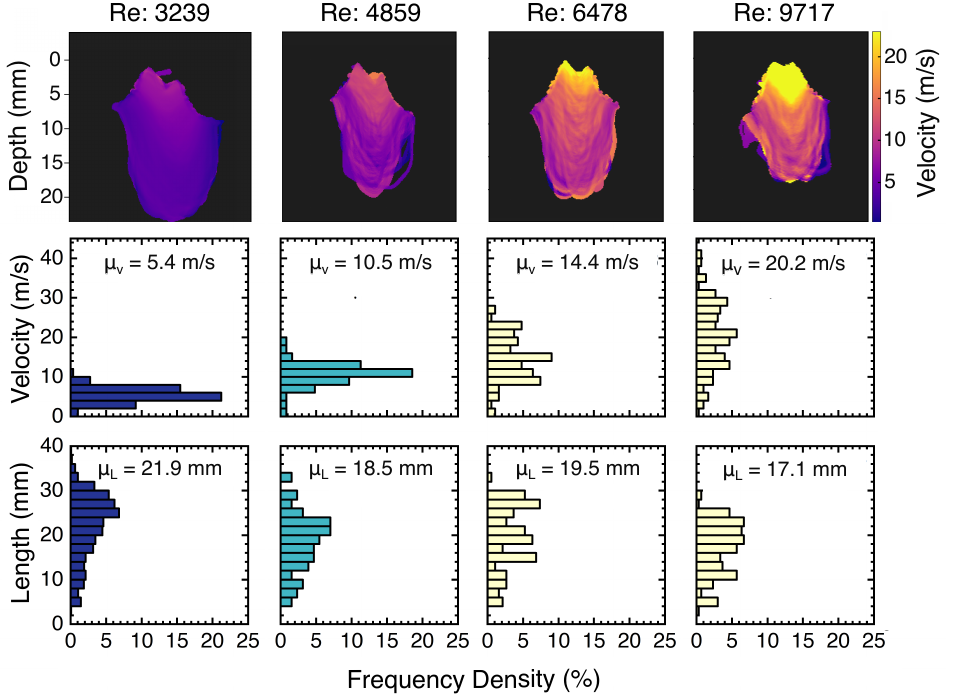}
\caption{
Statistical characterization of kinematic properties of the gliding arc discharge aggregated across many discharges and various Re showing broad velocity gradients at high Re with no relationship to average length.
}
\end{figure*}

Identification of these discharge behaviors provides a platform for more meaningful reactor optimization through a deeper understanding of arc behavior. For instance, work conducted by Wang et al. simulated that control over the type of arc formation in the reactor system can lead to an enhancement in energy efficiency.\cite{wang_gliding_2017} Establishing these discharge characteristics and their relationship to reactor configuration and flow environment provides a platform for more standardized reactor reporting and more direct pathways for optimization. As each mode is identified directly from the electrical waveform, these electrical classifications are well suited to real-time reactor monitoring, offering a practical route toward feedback-driven operation.

In general, literature has largely considered Modes A and B as independent associated events, wherein the occurrence of either Mode A or B doesn't influence the subsequent mode event. However, this study reveals that the modes discussed in this work are not freely variant but exhibit some coupling behaviors. Most prominently, because Mode C acts as a reactor shutdown, resulting in no plasma, the subsequent arc formed immediately after Mode C is always Mode A. Meanwhile the conditional probabilities of subsequent mode events from an initial event exhibit a greater likelihood for transitioning between initiation mode event types rather than cascades of a single modal event, as shown in \textbf{S1.5} over large datasets.

Mode B can be described as a semi-continuous state, in which cascades of Mode B events can occur in sequence, particularly at low flow rates where individual arc lifetimes are assumed to be longer.\cite{zhang_mode_2020} However, the periodic nature of Mode C and the Mode C-to-A coupling suppresses these cascading events. As such, the prevalence of Mode A and B is not purely random with this power supply but is instead shaped by the NST, and shows no trend in relation to Re; evaluation of Mode B prevalence from optical and electrical diagnostics can be found in \textbf{S1.5}. Establishing these electrical signatures is essential for the development of an electrical-only diagnostic tool that permits \textit{operando} reactor monitoring of arc dynamics, a capability that has been absent from the GAD literature. 

Additional insights can also be gathered from the optical analysis by tracking arc speeds, lengths, and penetration depths across various Re to assess travel speeds and typical length scales. Importantly, depth is defined as the furthest pixel distance from the minimum electrode gap reached by the tracked arc, while length is defined as the total centerline length of an arc, obtained by skeletonizing the plasma arc path, as described in the \textbf{S1.4}. Such analysis is provided in \textbf{Figure 5} wherein increasing Re produces faster moving arcs with broad velocity gradients, as demonstrated in the heatmaps and histograms. 

This indicates that the fastest moving portions of the gas occur at the narrowest electrode positions, where Mode A type discharges occur and slower moving gas velocities in regions outside the minimum electrode position where Mode B behaviors occur. Evidence for distinguishable velocities based on initiation type is provided in \textbf{S1.4}. In contrast, lower Re shows more uniform velocity gradients and a stronger clustering of velocity data. There is no apparent trend relating to arc length Re, as arc lengths vary widely across the experimental conditions tested. These insights provide useful quantitative information about arc behavior that electrical-only diagnostics cannot readily extract.

\subsection{Electrical Event Detection}
\begin{figure*}[t]
\centering
\includegraphics[width=0.9\linewidth, keepaspectratio]{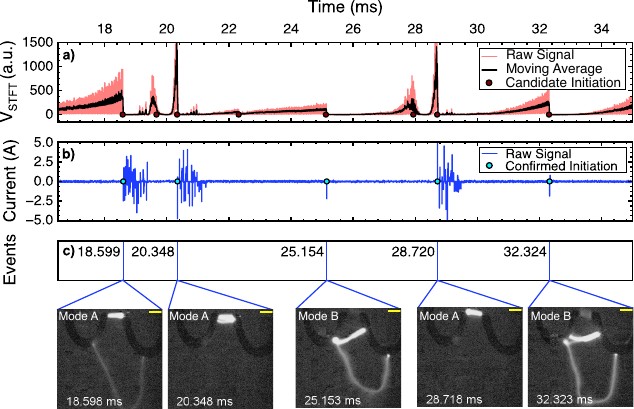}
\caption{
Electrical event detection aligns with high-speed imaging. (a) STFT of voltage identifies candidate initiation events spanning modes A–D. (b) Joint voltage and current analysis confirms mode A and B events. (c) Confirmed event timestamps with corresponding high-speed photography frames. Yellow scale bars represent 3.2 mm.
}
\end{figure*}

We subsequently developed computational methods to robustly track voltage drops and current spikes, identifying initiation events directly from electrical signals. To simplify the search for voltage drops, a short-time Fourier transform (STFT) was applied to the raw voltage signal, producing transformed signal shapes with well-defined drop-offs corresponding to decreases in the raw voltage; see \textbf{S1.6} for the derivation. A functional representation of the algorithm is provided in \textbf{S1.6}, wherein candidate initiation events are identified from voltage features alone, capturing all discharge events, Modes A–D, since each mode produces a voltage drop. The subset corresponding to initiation events, Modes A and B, are then identified by screening for current fluctuations at the time of each voltage drop. This two-stage approach reaffirms the necessity to evaluate both voltage and current simultaneously to differentiate arc events.

The developed tool was then validated against high-speed imaging in several representative GAD experiments, in which optical and electrical diagnostics were time-gated, as shown in \textbf{Figure 6}. Confirmed initiation events were flagged by the tool with cyan markers (\textbf{Figure 6b}), with the predicted time of discharge shown in \textbf{Figure 6c}. The provided images represent the earliest frame in which a new arc is observed. 

Characteristic peaks associated with Mode C are apparent from the V$_{STFT}$  peaks at $\approx$ 19 and 28 ms; these events are correctly excluded from the initiation screening via the lack of a current signal disruption. Similarly, a Mode D event appears near $\approx$ 22 ms, flagged as a candidate event in the V$_{STFT}$  waveform but also excluded from the initiation event subset due to the absence of a current disruption. 

Continued evidence for the validation of this tool across longer voltage–current waveforms as well as documentation regarding analysis thresholds is provided in \textbf{S1.6}. The developed tool, therefore, shows strong agreement between predicted initiation events and new arc formation observed in the optical diagnostics. 

\subsection{Mean Discharge Time as a Reactor Descriptor}
While classifying the four discrete modes of the GAD under study is valuable, the events associated with new arc formation, Modes A and B, are particularly worth evaluating for their relation to the electron density in the plasma arc and their presence across the literature, independent of power supply. The electron-density fluctuations accompanying these modes are critical to driving thermal and non-thermal chemistry.\cite{wang_gliding_2017,yang_numerical_2023}

\begin{figure*}[!t]
\centering
\includegraphics[width=0.9\linewidth, keepaspectratio]{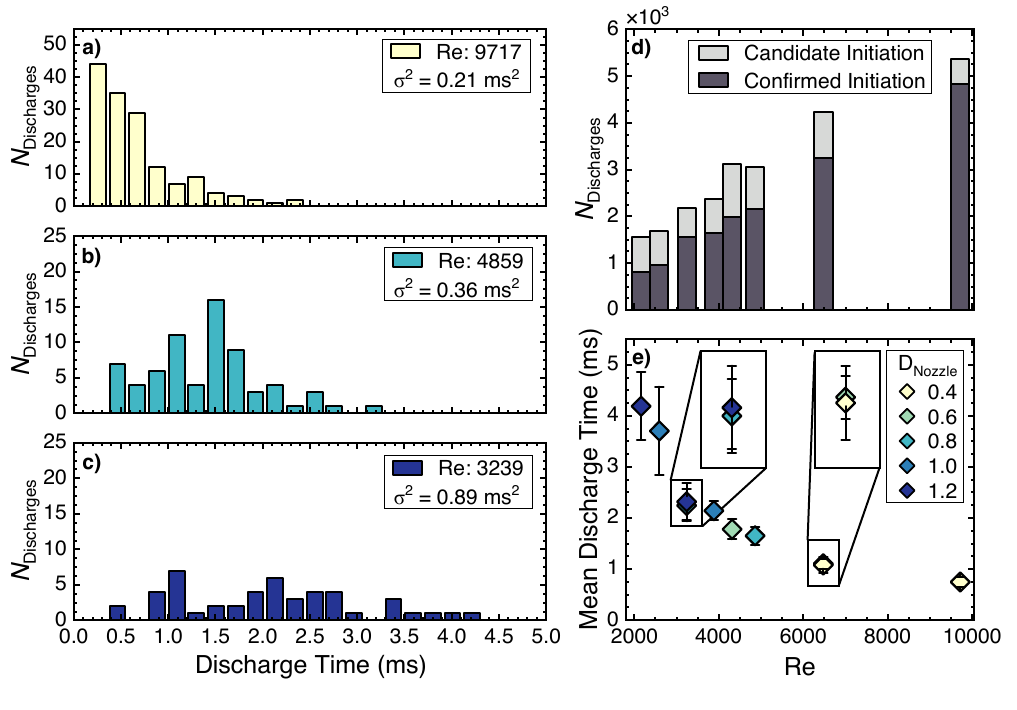}
\caption{
Statistical characterization of arc-formation intervals across Reynolds number and initiation vs all modal event tracking, demonstrating more frequent plasma events and less broad event distributions at higher Reynolds numbers. (a-c) Confirmed initiation event-time distributions from a single 100-ms voltage–current waveform at varying Reynolds numbers. (d) Collective discharge counts from 3-seconds worth of voltage-current waveforms from Mode A-D screening and Mode A-B screening at a fixed 13 mm electrode-to-nozzle distance. (e) Mean discharge time plotted as a function of Re from 3-seconds worth of voltage-current waveforms.
}
\end{figure*}

Consequently, the frequency with which a new arc forms becomes a characteristic parameter for describing reactor operation, as defined by both the reactor geometry and flow conditions. Using the electrical signatures described in the preceding sections, these initiation events can be explicitly tracked in the voltage and current waveforms, providing direct insight into the underlying arc dynamics.

Although modes A–D appear across all experimental conditions, the inherently heterogeneous nature of GADs makes a statistical analysis of bulk discharge behavior, evaluated over many waveforms, more informative than comparing individual discharges. Repeating this analysis across several waveforms reveals overall trends while reducing the influence of any single event misclassifications. 

Physically, a higher Reynolds number represents greater driving forces for arc propagation, which is expected to produce more frequent discharge events.\cite{kong_effect_2018,mutaf-yardimci_thermal_2000,zhang_mode_2018,zhang_mode_2020} \textbf{Figure 7a-c} shows discharge time distributions at three tested Reynolds numbers, each constructed from a single 100-ms electrical waveform of confirmed initiation events. This trend, which has been widely accepted qualitatively in the literature, is now demonstrated quantitatively. 

Specifically, at the highest Re, the distribution is narrow and clustered at low discharge times. As Re decreases, the distribution broadens, with longer and more variable discharge times as shown by the increase in variance, and a corresponding drop in the total number of events. This reveals that a single discharge time cannot fully characterize the arc dynamics of a GAD because these systems are inherently transient and heterogeneous. 

While no single value can fully capture the dynamic nature of discharge times, this analysis was extended from a single 100-ms waveform to thirty waveforms, for the characterization of collective trends. The results of this analysis are shown in \textbf{Figure 7d}, which separates contributions of all modal events, Modes A–D, as candidate initiation events, from the confirmed initiation events, Modes A and B. Here, a similarly increasing trend in total events with Re is observed, whereas the subset of confirmed events shows a much cleaner trend. 

From this analysis, the mean discharge time can be determined from each individual waveform distribution, \textbf{Figure 7a–c},and evaluated against Re. Importantly, the error bars on \textbf{Figure 7e} represent the deviation from the mean value aggregated across each waveform. An analytical comparison between optical and electrical discharge times is presented in \textbf{S1.5}.

The mean discharge time shows a decaying power-law relationship with increasing Re, with separate experiments at varying nozzle diameter and flow rate showing strong overlap at a shared Re, as shown in the insets; power-law fit parameters are provided in the \textbf{S1.5}. Additionally, the standard deviation in mean discharge time decreases over fivefold from the lowest to the highest Re tested, indicating that the discharge dynamics become substantially more stable and reproducible as flow conditions provide increased driving force for propagation. These results establish mean discharge time as a quantitative, flow-derived descriptor of reactor stability, providing the basis to directly evaluate its relationship to CO production rate.

\begin{figure}[h]
\centering
\includegraphics[width=0.9\linewidth, keepaspectratio]{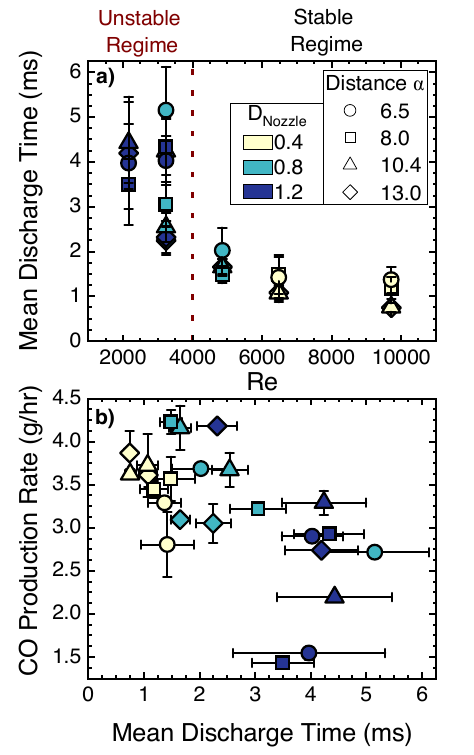}
\caption{
Mean discharge time stabilizes above Re $\approx$
 4000, and indicates reproducibility in the achieved production rate. (a) Mean discharge time across reactor geometries converges above Re $\approx$ 4000. (b) Production rate decreases with mean discharge time, with reduced variation at shorter discharge times.
}
\end{figure}

To understand how different reactor geometries influence the arc dynamics, various electrode-to-nozzle distances can be tested. Recall that the jet-to-plasma (JtP) overlap serves as an estimated upper bound for gas exposure to the plasma environment as the gas spreads outwards from the inlet nozzle. \textbf{Figure 8a} shows how the mean discharge time varies across different reactor geometries at a given Re.

Below Re $\approx$ 4000, discharge times are long and vary widely, while also demonstrating low CO production rate values as shown in \textbf{Figure 8b}, indicating that the instability in CO production rate is a direct result of instability in the discharge times under these conditions. Above this threshold, mean discharge times converge across the various geometries, with subtle variations. As such, stability in production rates across geometries reflects discharge times that have become less variable. Notably, this transitional Re between the stable and unstable regimes likely represents a threshold specific to the reactor geometry rather than a physically meaningful cutoff for operational stability.

A physical explanation for this variability lies in the non-linear travel paths arcs follow during propagation, wherein an arc can extend not only in a planar fashion along the electrodes but also off-center relative to the inlet nozzle. At low JtP overlap and high Re, the wide jet envelope and strong inertial forces ensure that all arcs following a non-ideal path are reliably carried toward extinction. In contrast, at high JtP overlap and low Re, the low inertial forces are concentrated in a narrow region, and arcs whose travel paths exit this region do not propagate outward reliably, resulting in the wide variations in discharge time observed at these conditions.

\textbf{Figure 8b} furthers this assessment by showing a clustering of short mean discharge times with high production rates, while configurations with variable discharge times produce lower and more variable CO production rates across different nozzle and distance $\alpha$ combinations. This establishes the need to consider both the arc dynamics at play and the inlet fluid flow path when designing and characterizing GADs, and provides mean discharge time as a quantitative descriptor that captures this behavior from electrical signals alone.

\section{Conclusion}\label{sec13}
This work establishes mean discharge time as an \textit{operando} descriptor of gliding arc dynamics, providing a quantitative link among fluid dynamics, reactor configuration, and arc dynamics that has historically been neglected in this promising electrified reactor technology. Using high-speed photography, four distinct arc behaviors characteristic of both the plasma and power supply were identified and linked to unique electrical signatures, establishing a new framework for mode classification in gliding arc systems. A computational method was developed to track these electrical signatures and validated against high-speed photography at 200,000 frames per second, providing an inter-frame interval roughly four times shorter than one period of the power source. 

High-speed imaging served solely to establish and validate the electrical signatures corresponding to each discharge mode. Once calibrated for a given reactor and gas composition, electrical measurements alone are sufficient to monitor and classify arc dynamics, without further reliance on optical diagnostics. Discharge time distributions were constructed across various operating conditions, and mean discharge time was found to follow a decaying power law scaling relationship with Re, and to be reproducible from different experimental conditions at matched Re.  

Variability in mean discharge time correlated directly with variability in the CO production rate, thereby providing a clear separation between stable and unstable operating regimes and establishing a direct link between arc dynamics, reactor geometry, and flow conditions. Increased CO$_2$ jet-to-plasma overlap failed to consistently improve CO$_2$  conversion suggesting that the arc cycling stability and arc formation frequency play a dominant role in optimizing reactor performance regardless of the chemistry leveraged. Mean discharge time is therefore proposed as an \textit{operando} diagnostic descriptor of gliding arc discharges, which should be leveraged to guide reactor design and optimization and can be extracted from the same voltage-current waveforms used to compute power consumption. Collectively, this work reports a low-power consumption of under 60 W, with a non-optimized reactor achieving a $\approx$ 5 \% single pass conversion and energy efficiencies exceeding 20 \% using a commercially available transformer as power supply to produce performance metrics directly comparable to existing literature without any catalyst integration.

Application of this framework to other systems requires joint optical and electrical diagnostics at high temporal resolutions to confirm the electrical signatures for the plasma modes with respect to the gas mixture and power supply under study, as both the discharge dynamics and power supply characteristics may affect these definitions.

\section{Methods}
The reactor and diagnostic equipment are shown schematically in Figure 1. Two diverging aluminum electrodes were formed from 2 mm diameter aluminum wire, bent to create an opening angle of 25° and a minimum inter-electrode gap of 3.2 mm with a bent arm length of 8 mm. The minimum inter-electrode gap was selected to keep the electrodes outside the inlet jet envelope across all experimental conditions tested, preventing the electrodes from influencing the fluid flow as discussed in the \textbf{S1.2}. The nozzle-to-electrode distance was adjustable between 6.5 and 13 mm, and commercially available 3D-printer nozzles, ranging in diameter from 0.4 to 1.2 mm in 0.2 mm increments, were threaded into the reactor. Pure CO$_{2}$ was injected at 1.0 and 1.5 SLPM with flows being regulated via an Alicat mass flow controller.

Plasma was generated using a neon sign transformer (15 kV, 40 mA, Model BNP-508) with floating high-voltage connections. Voltage was monitored using two high-voltage probes (Rigol RP1018H), current was measured using a Pearson coil (Model 4100), and all electrical signals were recorded using an oscilloscope (Tektronix MSO24). High-speed imaging was performed using a Phantom v2512 camera operating at 200,000 frames per second with a 4.53 $\mu$s exposure at 256×256 pixel resolution; camera imaging and oscilloscope measurements were synchronized using a pulse delay generator (Stanford Research Systems DG535). Gaseous products were tracked using a Hiden Analytical Mass Spectrometer (HPR-20 EGA) connected as a side stream to minimize sensitivity to pressure fluctuations during plasma operation. Each experimental condition was run for 7 minutes, with 100-ms electrical waveforms recorded at 30-second intervals to ensure power consumption values reached steady state.

Conversion was calculated from mass spectrometer signal intensities using a normalization across CO$_2$, CO, and O$_2$ peaks, with a correction term applied to account for gas dilution from the increase in gaseous moles produced by the reaction, see full derivation in the \textbf{S1.1}.\cite{wanten_plasma-based_2023}  

\begin{equation}
P_{\mathrm{plasma}}~(\mathrm{W}) = \frac{1}{ t}\times \int_0^t V(t) \times I(t) \, dt
\label{Power}
\end{equation}

In plasma reactors, there are two primary ways of describing power consumption. Plug-power, which describes the total power draw from the power supply, and plasma power, which is the power applied to generate a plasma after the transformation into high-voltage. By convention, plasma power (P$_{plasma}$) is most often reported to decouple transformer efficiency and the power supplied to the plasma, which drives chemical conversion.

Energy efficiency, energy cost, specific energy input, and CO production rate were calculated using standard definitions from the literature, as detailed in \textbf{S1.1}. Each experimental condition was measured in triplicate, and the error bars represent standard deviations across replicates. All data processing practices, threshold values, and calculation parameters for electrical and optical event detection have been summarized in the \textbf{S1.4-1.6}. All code used in this work has been provided as a publicly accessible database. 

\section{Acknowledgments}
We thank Breakthrough Energy Foundation for funding this research. Alexander Davis gratefully acknowledges support from the Ryan Fellowship and the International Institute for Nanotechnology at Northwestern University.

\section{Data Availability}
Experimental data, electrical and optical analysis scripts, and representative videos are housed in a public data repository for reproducibility. All code utilized in this work has been published to https://doi.org/10.5281/zenodo.21724305.

\section{Contributions}
Primary experiments were conducted by Alexander Davis. Charles Burton assisted in high-speed photography acquisition and processing. Stephanie Pecaut contributed to the electrical circuitry setup. Matt Hershey helped with the assembly of the experimental manifold and paper preparation. Linsey C. Seitz, Michelle M. Driscoll, and Dayne F. Swearer supervised and managed the research project. Dayne Swearer developed the experimental scope and secured scientific funding. Alexander Davis wrote the first draft, with input from all the authors. 

\section{Competing Interests}
The authors declare no competing interests.

\section{Supplementary Information}
Provided as an attached document.

\newpage
\clearpage

\bibliography{MethodsTrimmed}


\begin{thebibliography}{34}
\ifx \bisbn   \undefined \def \bisbn  #1{ISBN #1}\fi
\ifx \binits  \undefined \def \binits#1{#1}\fi
\ifx \bauthor  \undefined \def \bauthor#1{#1}\fi
\ifx \batitle  \undefined \def \batitle#1{#1}\fi
\ifx \bjtitle  \undefined \def \bjtitle#1{#1}\fi
\ifx \bvolume  \undefined \def \bvolume#1{\textbf{#1}}\fi
\ifx \byear  \undefined \def \byear#1{#1}\fi
\ifx \bissue  \undefined \def \bissue#1{#1}\fi
\ifx \bfpage  \undefined \def \bfpage#1{#1}\fi
\ifx \blpage  \undefined \def \blpage #1{#1}\fi
\ifx \burl  \undefined \def \burl#1{\textsf{#1}}\fi
\ifx \doiurl  \undefined \def \doiurl#1{\url{https://doi.org/#1}}\fi
\ifx \betal  \undefined \def \betal{\textit{et al.}}\fi
\ifx \binstitute  \undefined \def \binstitute#1{#1}\fi
\ifx \binstitutionaled  \undefined \def \binstitutionaled#1{#1}\fi
\ifx \bctitle  \undefined \def \bctitle#1{#1}\fi
\ifx \beditor  \undefined \def \beditor#1{#1}\fi
\ifx \bpublisher  \undefined \def \bpublisher#1{#1}\fi
\ifx \bbtitle  \undefined \def \bbtitle#1{#1}\fi
\ifx \bedition  \undefined \def \bedition#1{#1}\fi
\ifx \bseriesno  \undefined \def \bseriesno#1{#1}\fi
\ifx \blocation  \undefined \def \blocation#1{#1}\fi
\ifx \bsertitle  \undefined \def \bsertitle#1{#1}\fi
\ifx \bsnm \undefined \def \bsnm#1{#1}\fi
\ifx \bsuffix \undefined \def \bsuffix#1{#1}\fi
\ifx \bparticle \undefined \def \bparticle#1{#1}\fi
\ifx \barticle \undefined \def \barticle#1{#1}\fi
\bibcommenthead
\ifx \bconfdate \undefined \def \bconfdate #1{#1}\fi
\ifx \botherref \undefined \def \botherref #1{#1}\fi
\ifx \url \undefined \def \url#1{\textsf{#1}}\fi
\ifx \bchapter \undefined \def \bchapter#1{#1}\fi
\ifx \bbook \undefined \def \bbook#1{#1}\fi
\ifx \bcomment \undefined \def \bcomment#1{#1}\fi
\ifx \oauthor \undefined \def \oauthor#1{#1}\fi
\ifx \citeauthoryear \undefined \def \citeauthoryear#1{#1}\fi
\ifx \endbibitem  \undefined \def \endbibitem {}\fi
\ifx \bconflocation  \undefined \def \bconflocation#1{#1}\fi
\ifx \arxivurl  \undefined \def \arxivurl#1{\textsf{#1}}\fi
\csname PreBibitemsHook\endcsname

\bibitem[\protect\citeauthoryear{Bistline and Blanford}{2021}]{bistline_role_2021}
\begin{barticle}
\bauthor{\bsnm{Bistline}, \binits{J.E.T.}},
\bauthor{\bsnm{Blanford}, \binits{G.J.}}:
\batitle{The role of the power sector in net-zero energy systems}.
\bjtitle{Energy and Climate Change}
\bvolume{2},
\bfpage{100045}
(\byear{2021})
\doiurl{10.1016/j.egycc.2021.100045}
\end{barticle}
\endbibitem

\bibitem[\protect\citeauthoryear{Bouckaert et~al.}{2021}]{bouckaert_net_2021}
\begin{botherref}
\oauthor{\bsnm{Bouckaert}, \binits{S.}},
\oauthor{\bsnm{Pales}, \binits{A.F.}},
\oauthor{\bsnm{McGlade}, \binits{C.}},
\oauthor{\bsnm{Remme}, \binits{U.}},
\oauthor{\bsnm{Wanner}, \binits{B.}},
\oauthor{\bsnm{Varro}, \binits{L.}},
\oauthor{\bsnm{D'Ambrosio}, \binits{D.}},
\oauthor{\bsnm{Spencer}, \binits{T.}}:
Net {Zero} by 2050: {A} {Roadmap} for the {Global} {Energy} {Sector}
(2021)
\end{botherref}
\endbibitem

\bibitem[\protect\citeauthoryear{DeAngelo et~al.}{2021}]{deangelo_energy_2021}
\begin{barticle}
\bauthor{\bsnm{DeAngelo}, \binits{J.}},
\bauthor{\bsnm{Azevedo}, \binits{I.}},
\bauthor{\bsnm{Bistline}, \binits{J.}},
\bauthor{\bsnm{Clarke}, \binits{L.}},
\bauthor{\bsnm{Luderer}, \binits{G.}},
\bauthor{\bsnm{Byers}, \binits{E.}},
\bauthor{\bsnm{Davis}, \binits{S.J.}}:
\batitle{Energy systems in scenarios at net-zero {CO2} emissions}.
\bjtitle{Nature Communications}
\bvolume{12}(\bissue{1}),
\bfpage{6096}
(\byear{2021})
\doiurl{10.1038/s41467-021-26356-y}
\end{barticle}
\endbibitem

\bibitem[\protect\citeauthoryear{Mallapragada et~al.}{2023}]{mallapragada_decarbonization_2023}
\begin{barticle}
\bauthor{\bsnm{Mallapragada}, \binits{D.S.}},
\bauthor{\bsnm{Dvorkin}, \binits{Y.}},
\bauthor{\bsnm{Modestino}, \binits{M.A.}},
\bauthor{\bsnm{Esposito}, \binits{D.V.}},
\bauthor{\bsnm{Smith}, \binits{W.A.}},
\bauthor{\bsnm{Hodge}, \binits{B.-M.}},
\bauthor{\bsnm{Harold}, \binits{M.P.}},
\bauthor{\bsnm{Donnelly}, \binits{V.M.}},
\bauthor{\bsnm{Nuz}, \binits{A.}},
\bauthor{\bsnm{Bloomquist}, \binits{C.}},
\bauthor{\bsnm{Baker}, \binits{K.}},
\bauthor{\bsnm{Grabow}, \binits{L.C.}},
\bauthor{\bsnm{Yan}, \binits{Y.}},
\bauthor{\bsnm{Rajput}, \binits{N.N.}},
\bauthor{\bsnm{Hartman}, \binits{R.L.}},
\bauthor{\bsnm{Biddinger}, \binits{E.J.}},
\bauthor{\bsnm{Aydil}, \binits{E.S.}},
\bauthor{\bsnm{Taylor}, \binits{A.D.}}:
\batitle{Decarbonization of the chemical industry through electrification: {Barriers} and opportunities}.
\bjtitle{Joule}
\bvolume{7}(\bissue{1}),
\bfpage{23}--\blpage{41}
(\byear{2023})
\doiurl{10.1016/j.joule.2022.12.008}
\end{barticle}
\endbibitem

\bibitem[\protect\citeauthoryear{Cresko et~al.}{2022}]{cresko_us_2022}
\begin{botherref}
\oauthor{\bsnm{Cresko}, \binits{J.}},
\oauthor{\bsnm{Rightor}, \binits{E.}},
\oauthor{\bsnm{Carpenter}, \binits{A.}},
\oauthor{\bsnm{Peretti}, \binits{K.}},
\oauthor{\bsnm{Elliott}, \binits{N.}},
\oauthor{\bsnm{Nimbalkar}, \binits{S.}},
\oauthor{\bsnm{Morrow~Iii}, \binits{W.}},
\oauthor{\bsnm{Hasanbeigi}, \binits{A.}},
\oauthor{\bsnm{Hedman}, \binits{B.}},
\oauthor{\bsnm{Supekar}, \binits{S.}},
\oauthor{\bsnm{McMillan}, \binits{C.}},
\oauthor{\bsnm{Hoffmeister}, \binits{A.}},
\oauthor{\bsnm{Whitlock}, \binits{A.}},
\oauthor{\bsnm{Igogo}, \binits{T.}},
\oauthor{\bsnm{Walzberg}, \binits{J.}},
\oauthor{\bsnm{D'Alessandro}, \binits{C.}},
\oauthor{\bsnm{Anderson}, \binits{S.}},
\oauthor{\bsnm{Atnoorkar}, \binits{S.}},
\oauthor{\bsnm{Upsani}, \binits{S.}},
\oauthor{\bsnm{King}, \binits{P.}},
\oauthor{\bsnm{Grgich}, \binits{J.}},
\oauthor{\bsnm{Ovard}, \binits{L.}},
\oauthor{\bsnm{Foist}, \binits{R.}},
\oauthor{\bsnm{Conner}, \binits{A.}},
\oauthor{\bsnm{Meshek}, \binits{M.}},
\oauthor{\bsnm{Hicks}, \binits{A.}},
\oauthor{\bsnm{Dollinger}, \binits{C.}},
\oauthor{\bsnm{Liddell}, \binits{H.}}:
U.{S}. {Department} of {Energy}’s {Industrial} {Decarbonization} {Roadmap}.
Technical Report DOE/EE--2635, 1961393
(September 2022).
\doiurl{10.2172/1961393}
\end{botherref}
\endbibitem

\bibitem[\protect\citeauthoryear{Osorio-Tejada et~al.}{2024}]{osorio-tejada_co2_2024}
\begin{barticle}
\bauthor{\bsnm{Osorio-Tejada}, \binits{J.}},
\bauthor{\bsnm{Escriba-Gelonch}, \binits{M.}},
\bauthor{\bsnm{Vertongen}, \binits{R.}},
\bauthor{\bsnm{Bogaerts}, \binits{A.}},
\bauthor{\bsnm{Hessel}, \binits{V.}}:
\batitle{{CO2} conversion to {CO} via plasma and electrolysis: a techno-economic and energy cost analysis}.
\bjtitle{Energy \& Environmental Science}
\bvolume{17}(\bissue{16}),
\bfpage{5833}--\blpage{5853}
(\byear{2024})
\doiurl{10.1039/D4EE00164H}
\end{barticle}
\endbibitem

\bibitem[\protect\citeauthoryear{George et~al.}{2021}]{george_review_2021}
\begin{barticle}
\bauthor{\bsnm{George}, \binits{A.}},
\bauthor{\bsnm{Shen}, \binits{B.}},
\bauthor{\bsnm{Craven}, \binits{M.}},
\bauthor{\bsnm{Wang}, \binits{Y.}},
\bauthor{\bsnm{Kang}, \binits{D.}},
\bauthor{\bsnm{Wu}, \binits{C.}},
\bauthor{\bsnm{Tu}, \binits{X.}}:
\batitle{A {Review} of {Non}-{Thermal} {Plasma} {Technology}: {A} novel solution for {CO2} conversion and utilization}.
\bjtitle{Renewable and Sustainable Energy Reviews}
\bvolume{135},
\bfpage{109702}
(\byear{2021})
\doiurl{10.1016/j.rser.2020.109702}
\end{barticle}
\endbibitem

\bibitem[\protect\citeauthoryear{Snoeckx and Bogaerts}{2017}]{snoeckx_plasma_2017}
\begin{barticle}
\bauthor{\bsnm{Snoeckx}, \binits{R.}},
\bauthor{\bsnm{Bogaerts}, \binits{A.}}:
\batitle{Plasma technology – a novel solution for {CO2} conversion?}
\bjtitle{Chemical Society Reviews}
\bvolume{46}(\bissue{19}),
\bfpage{5805}--\blpage{5863}
(\byear{2017})
\doiurl{10.1039/C6CS00066E}
\end{barticle}
\endbibitem

\bibitem[\protect\citeauthoryear{Fridman}{2008}]{fridman_plasma_2008}
\begin{bbook}
\bauthor{\bsnm{Fridman}, \binits{A.A.}}:
\bbtitle{Plasma Chemistry}.
\bpublisher{Cambridge University Press},
\blocation{Cambridge ;}
(\byear{2008})
\end{bbook}
\endbibitem

\bibitem[\protect\citeauthoryear{Adamovich et~al.}{2022}]{adamovich_2022_2022}
\begin{barticle}
\bauthor{\bsnm{Adamovich}, \binits{I.}},
\bauthor{\bsnm{Agarwal}, \binits{S.}},
\bauthor{\bsnm{Ahedo}, \binits{E.}},
\bauthor{\bsnm{Alves}, \binits{L.L.}},
\bauthor{\bsnm{Baalrud}, \binits{S.}},
\bauthor{\bsnm{Babaeva}, \binits{N.}},
\bauthor{\bsnm{Bogaerts}, \binits{A.}},
\bauthor{\bsnm{Bourdon}, \binits{A.}},
\bauthor{\bsnm{Bruggeman}, \binits{P.J.}},
\bauthor{\bsnm{Canal}, \binits{C.}},
\bauthor{\bsnm{Choi}, \binits{E.H.}},
\bauthor{\bsnm{Coulombe}, \binits{S.}},
\bauthor{\bsnm{Donkó}, \binits{Z.}},
\bauthor{\bsnm{Graves}, \binits{D.B.}},
\bauthor{\bsnm{Hamaguchi}, \binits{S.}},
\bauthor{\bsnm{Hegemann}, \binits{D.}},
\bauthor{\bsnm{Hori}, \binits{M.}},
\bauthor{\bsnm{Kim}, \binits{H.-H.}},
\bauthor{\bsnm{Kroesen}, \binits{G.M.W.}},
\bauthor{\bsnm{Kushner}, \binits{M.J.}},
\bauthor{\bsnm{Laricchiuta}, \binits{A.}},
\bauthor{\bsnm{Li}, \binits{X.}},
\bauthor{\bsnm{Magin}, \binits{T.E.}},
\bauthor{\bsnm{Mededovic~Thagard}, \binits{S.}},
\bauthor{\bsnm{Miller}, \binits{V.}},
\bauthor{\bsnm{Murphy}, \binits{A.B.}},
\bauthor{\bsnm{Oehrlein}, \binits{G.S.}},
\bauthor{\bsnm{Puac}, \binits{N.}},
\bauthor{\bsnm{Sankaran}, \binits{R.M.}},
\bauthor{\bsnm{Samukawa}, \binits{S.}},
\bauthor{\bsnm{Shiratani}, \binits{M.}},
\bauthor{\bsnm{Šimek}, \binits{M.}},
\bauthor{\bsnm{Tarasenko}, \binits{N.}},
\bauthor{\bsnm{Terashima}, \binits{K.}},
\bauthor{\bsnm{Thomas~Jr}, \binits{E.}},
\bauthor{\bsnm{Trieschmann}, \binits{J.}},
\bauthor{\bsnm{Tsikata}, \binits{S.}},
\bauthor{\bsnm{Turner}, \binits{M.M.}},
\bauthor{\bsnm{Walt}, \binits{I.J.}},
\bauthor{\bsnm{Sanden}, \binits{M.C.M.}},
\bauthor{\bsnm{Woedtke}, \binits{T.}}:
\batitle{The 2022 {Plasma} {Roadmap}: low temperature plasma science and technology}.
\bjtitle{Journal of Physics D: Applied Physics}
\bvolume{55}(\bissue{37}),
\bfpage{373001}
(\byear{2022})
\doiurl{10.1088/1361-6463/ac5e1c}
\end{barticle}
\endbibitem

\bibitem[\protect\citeauthoryear{O’Modhrain et~al.}{2024}]{omodhrain_upscaling_2024}
\begin{barticle}
\bauthor{\bsnm{O’Modhrain}, \binits{C.}},
\bauthor{\bsnm{Trenchev}, \binits{G.}},
\bauthor{\bsnm{Gorbanev}, \binits{Y.}},
\bauthor{\bsnm{Bogaerts}, \binits{A.}}:
\batitle{Upscaling {Plasma}-{Based} {CO}$_{\textrm{2}}$ {Conversion}: {Case} {Study} of a {Multi}-{Reactor} {Gliding} {Arc} {Plasmatron}}.
\bjtitle{ACS Engineering Au}
\bvolume{4}(\bissue{3}),
\bfpage{333}--\blpage{344}
(\byear{2024})
\doiurl{10.1021/acsengineeringau.3c00067}
\end{barticle}
\endbibitem

\bibitem[\protect\citeauthoryear{Wang et~al.}{2017}]{wang_gliding_2017}
\begin{barticle}
\bauthor{\bsnm{Wang}, \binits{W.}},
\bauthor{\bsnm{Mei}, \binits{D.}},
\bauthor{\bsnm{Tu}, \binits{X.}},
\bauthor{\bsnm{Bogaerts}, \binits{A.}}:
\batitle{Gliding arc plasma for {CO2} conversion: {Better} insights by a combined experimental and modelling approach}.
\bjtitle{Chemical Engineering Journal}
\bvolume{330},
\bfpage{11}--\blpage{25}
(\byear{2017})
\doiurl{10.1016/j.cej.2017.07.133}
\end{barticle}
\endbibitem

\bibitem[\protect\citeauthoryear{Czernichowski et~al.}{1996}]{czernichowski_spectral_1996}
\begin{barticle}
\bauthor{\bsnm{Czernichowski}, \binits{A.}},
\bauthor{\bsnm{Nassar}, \binits{H.}},
\bauthor{\bsnm{Ranaivosoloarimanana}, \binits{A.}},
\bauthor{\bsnm{Fridman}, \binits{A.A.}},
\bauthor{\bsnm{Simek}, \binits{M.}},
\bauthor{\bsnm{Musiol}, \binits{K.}},
\bauthor{\bsnm{Pawelec}, \binits{E.}},
\bauthor{\bsnm{Dittrichova}, \binits{L.}}:
\batitle{Spectral and {Electrical} {Diagnostics} of {Gliding} {Arc}}.
\bjtitle{Acta Physica Polonica A}
\bvolume{89}(\bissue{5-6}),
\bfpage{595}--\blpage{603}
(\byear{1996})
\doiurl{10.12693/APhysPolA.89.595}
\end{barticle}
\endbibitem

\bibitem[\protect\citeauthoryear{Wang et~al.}{2017}]{wang_nitrogen_2017}
\begin{barticle}
\bauthor{\bsnm{Wang}, \binits{W.}},
\bauthor{\bsnm{Patil}, \binits{B.}},
\bauthor{\bsnm{Heijkers}, \binits{S.}},
\bauthor{\bsnm{Hessel}, \binits{V.}},
\bauthor{\bsnm{Bogaerts}, \binits{A.}}:
\batitle{Nitrogen {Fixation} by {Gliding} {Arc} {Plasma}: {Better} {Insight} by {Chemical} {Kinetics} {Modelling}}.
\bjtitle{ChemSusChem}
\bvolume{10}(\bissue{10}),
\bfpage{2145}--\blpage{2157}
(\byear{2017})
\doiurl{10.1002/cssc.201700095}
\end{barticle}
\endbibitem

\bibitem[\protect\citeauthoryear{van Raak et~al.}{}]{van_raak_numbering_nodate}
\begin{botherref}
\oauthor{\bsnm{Raak}, \binits{T.}},
\oauthor{\bsnm{Bogaard}, \binits{H.}},
\oauthor{\bsnm{De~Felice}, \binits{G.}},
\oauthor{\bsnm{Emmery}, \binits{D.}},
\oauthor{\bsnm{Gallucci}, \binits{F.}},
\oauthor{\bsnm{Li}, \binits{S.}}:
Numbering up and sizing up gliding arc reactors to enhance the plasma-based synthesis of {NOx}.
Catalysis Science \& Technology
\textbf{14}(18),
5405--5421
\doiurl{10.1039/d4cy00655k}
\end{botherref}
\endbibitem

\bibitem[\protect\citeauthoryear{Gong et~al.}{2020}]{gong_decomposition_2020}
\begin{barticle}
\bauthor{\bsnm{Gong}, \binits{X.}},
\bauthor{\bsnm{Lin}, \binits{Y.}},
\bauthor{\bsnm{Li}, \binits{X.}},
\bauthor{\bsnm{Wu}, \binits{A.}},
\bauthor{\bsnm{Zhang}, \binits{H.}},
\bauthor{\bsnm{Yan}, \binits{J.}},
\bauthor{\bsnm{Du}, \binits{C.}}:
\batitle{Decomposition of volatile organic compounds using gliding arc discharge plasma}.
\bjtitle{Journal of the Air \& Waste Management Association}
\bvolume{70}(\bissue{2}),
\bfpage{138}--\blpage{157}
(\byear{2020})
\doiurl{10.1080/10962247.2019.1698476}
\end{barticle}
\endbibitem

\bibitem[\protect\citeauthoryear{Hameedl and Kadhem}{2020}]{hameedl_gliding_2020}
\begin{barticle}
\bauthor{\bsnm{Hameedl}, \binits{T.A.}},
\bauthor{\bsnm{Kadhem}, \binits{S.J.}}:
\batitle{Gliding arc discharge for water treatment}.
\bjtitle{IOP Conference Series: Materials Science and Engineering}
\bvolume{757}(\bissue{1}),
\bfpage{012045}
(\byear{2020})
\doiurl{10.1088/1757-899X/757/1/012045}
\end{barticle}
\endbibitem

\bibitem[\protect\citeauthoryear{Potočňáková et~al.}{2017}]{potocnakova_experimental_2017}
\begin{barticle}
\bauthor{\bsnm{Potočňáková}, \binits{L.}},
\bauthor{\bsnm{Šperka}, \binits{J.}},
\bauthor{\bsnm{Zikán}, \binits{P.}},
\bauthor{\bsnm{Van~Loon}, \binits{J.J.W.A.}},
\bauthor{\bsnm{Beckers}, \binits{J.}},
\bauthor{\bsnm{Kudrle}, \binits{V.}}:
\batitle{Experimental study of gliding arc plasma channel motion: buoyancy and gas flow phenomena under normal and hypergravity conditions}.
\bjtitle{Plasma Sources Science and Technology}
\bvolume{26}(\bissue{4}),
\bfpage{045014}
(\byear{2017})
\doiurl{10.1088/1361-6595/aa5ee8}
\end{barticle}
\endbibitem

\bibitem[\protect\citeauthoryear{Mutaf-Yardimci et~al.}{2000}]{mutaf-yardimci_thermal_2000}
\begin{barticle}
\bauthor{\bsnm{Mutaf-Yardimci}, \binits{O.}},
\bauthor{\bsnm{Saveliev}, \binits{A.V.}},
\bauthor{\bsnm{Fridman}, \binits{A.A.}},
\bauthor{\bsnm{Kennedy}, \binits{L.A.}}:
\batitle{Thermal and nonthermal regimes of gliding arc discharge in air flow}.
\bjtitle{Journal of Applied Physics}
\bvolume{87}(\bissue{4}),
\bfpage{1632}--\blpage{1641}
(\byear{2000})
\doiurl{10.1063/1.372071}
\end{barticle}
\endbibitem

\bibitem[\protect\citeauthoryear{Bryssinck et~al.}{2025}]{bryssinck_performance_2025}
\begin{barticle}
\bauthor{\bsnm{Bryssinck}, \binits{R.}},
\bauthor{\bsnm{Smith}, \binits{G.J.}},
\bauthor{\bsnm{O'Modhrain}, \binits{C.}},
\bauthor{\bsnm{Assche}, \binits{T.V.}},
\bauthor{\bsnm{Trenchev}, \binits{G.}},
\bauthor{\bsnm{Bogaerts}, \binits{A.}}:
\batitle{Performance of a gliding arc plasmatron pilot reactor with an integrated carbon bed and recirculation for upscaled {CO2} conversion}.
\bjtitle{Reaction Chemistry \& Engineering}
\bvolume{10}(\bissue{8}),
\bfpage{1910}--\blpage{1923}
(\byear{2025})
\doiurl{10.1039/D5RE00190K}
\end{barticle}
\endbibitem

\bibitem[\protect\citeauthoryear{Luo et~al.}{2025}]{luo_stabilizing_2025}
\begin{barticle}
\bauthor{\bsnm{Luo}, \binits{Z.}},
\bauthor{\bsnm{Sun}, \binits{H.}},
\bauthor{\bsnm{Wu}, \binits{Y.}},
\bauthor{\bsnm{Rong}, \binits{M.}},
\bauthor{\bsnm{Liu}, \binits{Z.}}:
\batitle{Stabilizing {Discharge} and {Suppressing} {Recombination}: {Advancing} {Carbon} {Conversion} {With} {Nozzle}-{Type} {Gliding} {Arc} {Discharge}}.
\bjtitle{Plasma Processes and Polymers}
\bvolume{22}(\bissue{11}),
\bfpage{70079}
(\byear{2025})
\doiurl{10.1002/ppap.70079}
\end{barticle}
\endbibitem

\bibitem[\protect\citeauthoryear{Pope}{2000}]{pope_turbulent_2000}
\begin{bbook}
\bauthor{\bsnm{Pope}, \binits{S.B.}}:
\bbtitle{Turbulent Flows}.
\bpublisher{Cambridge University Press},
\blocation{Cambridge}
(\byear{2000})
\end{bbook}
\endbibitem

\bibitem[\protect\citeauthoryear{Li et~al.}{2019}]{li_plasma-assisted_2019}
\begin{barticle}
\bauthor{\bsnm{Li}, \binits{L.}},
\bauthor{\bsnm{Zhang}, \binits{H.}},
\bauthor{\bsnm{Li}, \binits{X.}},
\bauthor{\bsnm{Kong}, \binits{X.}},
\bauthor{\bsnm{Xu}, \binits{R.}},
\bauthor{\bsnm{Tay}, \binits{K.}},
\bauthor{\bsnm{Tu}, \binits{X.}}:
\batitle{Plasma-assisted {CO2} conversion in a gliding arc discharge: {Improving} performance by optimizing the reactor design}.
\bjtitle{Journal of CO2 Utilization}
\bvolume{29},
\bfpage{296}--\blpage{303}
(\byear{2019})
\doiurl{10.1016/j.jcou.2018.12.019}
\end{barticle}
\endbibitem

\bibitem[\protect\citeauthoryear{Zhang and Luo}{2018}]{zhang_mode_2018}
\begin{bchapter}
\bauthor{\bsnm{Zhang}, \binits{R.}},
\bauthor{\bsnm{Luo}, \binits{G.}}:
\bctitle{The mode of gliding arc discharge and its characteristics}.
In: \bbtitle{2018 12th {International} {Conference} on the {Properties} and {Applications} of {Dielectric} {Materials} ({ICPADM})},
pp. \bfpage{305}--\blpage{310}
(\byear{2018}).
\doiurl{10.1109/ICPADM.2018.8401270}
\end{bchapter}
\endbibitem

\bibitem[\protect\citeauthoryear{Zhang et~al.}{2020}]{zhang_mode_2020}
\begin{barticle}
\bauthor{\bsnm{Zhang}, \binits{R.}},
\bauthor{\bsnm{Huang}, \binits{H.}},
\bauthor{\bsnm{Yang}, \binits{T.}}:
\batitle{Mode transition induced by back-breakdown of the gliding arc and its influence factors}.
\bjtitle{High Voltage}
\bvolume{5}(\bissue{3}),
\bfpage{306}--\blpage{312}
(\byear{2020})
\doiurl{10.1049/hve.2019.0162}
\end{barticle}
\endbibitem

\bibitem[\protect\citeauthoryear{Chen et~al.}{2021}]{chen_spatiotemporally_2021}
\begin{barticle}
\bauthor{\bsnm{Chen}, \binits{Z.}},
\bauthor{\bsnm{Yu}, \binits{J.}},
\bauthor{\bsnm{Cheng}, \binits{W.}},
\bauthor{\bsnm{Jiang}, \binits{Y.}},
\bauthor{\bsnm{Jiang}, \binits{L.}},
\bauthor{\bsnm{Tian}, \binits{Y.}},
\bauthor{\bsnm{Zhang}, \binits{L.}}:
\batitle{Spatiotemporally resolved characteristics of {AC} three-dimensional rotating gliding arc at atmospheric pressure}.
\bjtitle{Journal of Physics D: Applied Physics}
\bvolume{54}(\bissue{22}),
\bfpage{225203}
(\byear{2021})
\doiurl{10.1088/1361-6463/abea3a}
\end{barticle}
\endbibitem

\bibitem[\protect\citeauthoryear{Ma et~al.}{2025}]{ma_insight_2025}
\begin{barticle}
\bauthor{\bsnm{Ma}, \binits{X.}},
\bauthor{\bsnm{Liu}, \binits{K.}},
\bauthor{\bsnm{Liao}, \binits{H.}},
\bauthor{\bsnm{Zhou}, \binits{X.}}:
\batitle{Insight into the wire-plate gliding arc discharge supplied by a pulse-modulated {AC} power: {Superior} stability and discharge characteristics}.
\bjtitle{Physics of Plasmas}
\bvolume{32}(\bissue{5}),
\bfpage{053503}
(\byear{2025})
\doiurl{10.1063/5.0253496}
\end{barticle}
\endbibitem

\bibitem[\protect\citeauthoryear{Wang et~al.}{2022}]{wang_hydroxyl_2022}
\begin{barticle}
\bauthor{\bsnm{Wang}, \binits{Z.}},
\bauthor{\bsnm{Stamatoglou}, \binits{P.}},
\bauthor{\bsnm{Kong}, \binits{C.}},
\bauthor{\bsnm{Gao}, \binits{J.}},
\bauthor{\bsnm{Bao}, \binits{Y.}},
\bauthor{\bsnm{Aldén}, \binits{M.}},
\bauthor{\bsnm{Ehn}, \binits{A.}},
\bauthor{\bsnm{Richter}, \binits{M.}}:
\batitle{Hydroxyl radical dynamics in a gliding arc discharge using high-speed {PLIF} imaging}.
\bjtitle{Plasma Research Express}
\bvolume{4}(\bissue{2}),
\bfpage{025007}
(\byear{2022})
\doiurl{10.1088/2516-1067/ac76a4}
\end{barticle}
\endbibitem

\bibitem[\protect\citeauthoryear{Choi et~al.}{2025}]{choi_characterization_2025}
\begin{barticle}
\bauthor{\bsnm{Choi}, \binits{J.}},
\bauthor{\bsnm{Choi}, \binits{S.}},
\bauthor{\bsnm{Song}, \binits{Y.-H.}},
\bauthor{\bsnm{Lee}, \binits{D.H.}}:
\batitle{Characterization of gliding arc discharge using {H2}/{Ar} gas mixture}.
\bjtitle{International Journal of Hydrogen Energy}
\bvolume{106},
\bfpage{888}--\blpage{895}
(\byear{2025})
\doiurl{10.1016/j.ijhydene.2025.01.394}
\end{barticle}
\endbibitem

\bibitem[\protect\citeauthoryear{Zhu et~al.}{2014}]{zhu_dynamics_2014}
\begin{barticle}
\bauthor{\bsnm{Zhu}, \binits{J.}},
\bauthor{\bsnm{Sun}, \binits{Z.}},
\bauthor{\bsnm{Li}, \binits{Z.}},
\bauthor{\bsnm{Ehn}, \binits{A.}},
\bauthor{\bsnm{Aldén}, \binits{M.}},
\bauthor{\bsnm{Salewski}, \binits{M.}},
\bauthor{\bsnm{Leipold}, \binits{F.}},
\bauthor{\bsnm{Kusano}, \binits{Y.}}:
\batitle{Dynamics, {OH} distributions and {UV} emission of a gliding arc at various flow-rates investigated by optical measurements}.
\bjtitle{Journal of Physics D: Applied Physics}
\bvolume{47}(\bissue{29}),
\bfpage{295203}
(\byear{2014})
\doiurl{10.1088/0022-3727/47/29/295203}
\end{barticle}
\endbibitem

\bibitem[\protect\citeauthoryear{Bourlet et~al.}{2022}]{bourlet_numerical_2022}
\begin{bchapter}
\bauthor{\bsnm{Bourlet}, \binits{A.}},
\bauthor{\bsnm{Labaune}, \binits{J.}},
\bauthor{\bsnm{Tholin}, \binits{F.}},
\bauthor{\bsnm{Vincent}, \binits{A.}},
\bauthor{\bsnm{Pechereau}, \binits{F.}},
\bauthor{\bsnm{Laux}, \binits{C.O.}}:
\bctitle{Numerical model of restrikes in {DC} gliding arc discharges}.
In: \bbtitle{{AIAA} {SCITECH} 2022 {Forum}}.
\bpublisher{American Institute of Aeronautics and Astronautics},
\blocation{San Diego, CA \& Virtual}
(\byear{2022}).
\doiurl{10.2514/6.2022-0831}
\end{bchapter}
\endbibitem

\bibitem[\protect\citeauthoryear{Yang et~al.}{2023}]{yang_numerical_2023}
\begin{bchapter}
\bauthor{\bsnm{Yang}, \binits{M.}},
\bauthor{\bsnm{Wang}, \binits{Z.}},
\bauthor{\bsnm{Zhang}, \binits{W.}},
\bauthor{\bsnm{Zhang}, \binits{J.}},
\bauthor{\bsnm{Zhang}, \binits{H.}},
\bauthor{\bsnm{Ji}, \binits{Y.}}:
\bctitle{Numerical {Simulation} of {Gliding} {Arc} {Plasma} {Motion} {Characteristics}}.
In: \beditor{\bsnm{Dai}, \binits{D.}},
\beditor{\bsnm{Zhang}, \binits{C.}},
\beditor{\bsnm{Fang}, \binits{Z.}},
\beditor{\bsnm{Lu}, \binits{X.}} (eds.)
\bbtitle{Proceedings of the 4th {International} {Symposium} on {Plasma} and {Energy} {Conversion}},
pp. \bfpage{339}--\blpage{351}.
\bpublisher{Springer},
\blocation{Singapore}
(\byear{2023}).
\doiurl{10.1007/978-981-99-1576-7_32}
\end{bchapter}
\endbibitem

\bibitem[\protect\citeauthoryear{Kong et~al.}{2018}]{kong_effect_2018}
\begin{barticle}
\bauthor{\bsnm{Kong}, \binits{C.}},
\bauthor{\bsnm{Gao}, \binits{J.}},
\bauthor{\bsnm{Zhu}, \binits{J.}},
\bauthor{\bsnm{Ehn}, \binits{A.}},
\bauthor{\bsnm{Aldén}, \binits{M.}},
\bauthor{\bsnm{Li}, \binits{Z.}}:
\batitle{Effect of turbulent flow on an atmospheric-pressure {AC} powered gliding arc discharge}.
\bjtitle{Journal of Applied Physics}
\bvolume{123}(\bissue{22}),
\bfpage{223302}
(\byear{2018})
\doiurl{10.1063/1.5026703}
\end{barticle}
\endbibitem

\bibitem[\protect\citeauthoryear{Wanten et~al.}{2023}]{wanten_plasma-based_2023}
\begin{barticle}
\bauthor{\bsnm{Wanten}, \binits{B.}},
\bauthor{\bsnm{Vertongen}, \binits{R.}},
\bauthor{\bsnm{De~Meyer}, \binits{R.}},
\bauthor{\bsnm{Bogaerts}, \binits{A.}}:
\batitle{Plasma-based {CO2} conversion: {How} to correctly analyze the performance?}
\bjtitle{Journal of Energy Chemistry}
\bvolume{86},
\bfpage{180}--\blpage{196}
(\byear{2023})
\doiurl{10.1016/j.jechem.2023.07.005}
\end{barticle}
\endbibitem

\end{thebibliography}



\begin{thebibliography}{17}
\ifx \bisbn   \undefined \def \bisbn  #1{ISBN #1}\fi
\ifx \binits  \undefined \def \binits#1{#1}\fi
\ifx \bauthor  \undefined \def \bauthor#1{#1}\fi
\ifx \batitle  \undefined \def \batitle#1{#1}\fi
\ifx \bjtitle  \undefined \def \bjtitle#1{#1}\fi
\ifx \bvolume  \undefined \def \bvolume#1{\textbf{#1}}\fi
\ifx \byear  \undefined \def \byear#1{#1}\fi
\ifx \bissue  \undefined \def \bissue#1{#1}\fi
\ifx \bfpage  \undefined \def \bfpage#1{#1}\fi
\ifx \blpage  \undefined \def \blpage #1{#1}\fi
\ifx \burl  \undefined \def \burl#1{\textsf{#1}}\fi
\ifx \doiurl  \undefined \def \doiurl#1{\url{https://doi.org/#1}}\fi
\ifx \betal  \undefined \def \betal{\textit{et al.}}\fi
\ifx \binstitute  \undefined \def \binstitute#1{#1}\fi
\ifx \binstitutionaled  \undefined \def \binstitutionaled#1{#1}\fi
\ifx \bctitle  \undefined \def \bctitle#1{#1}\fi
\ifx \beditor  \undefined \def \beditor#1{#1}\fi
\ifx \bpublisher  \undefined \def \bpublisher#1{#1}\fi
\ifx \bbtitle  \undefined \def \bbtitle#1{#1}\fi
\ifx \bedition  \undefined \def \bedition#1{#1}\fi
\ifx \bseriesno  \undefined \def \bseriesno#1{#1}\fi
\ifx \blocation  \undefined \def \blocation#1{#1}\fi
\ifx \bsertitle  \undefined \def \bsertitle#1{#1}\fi
\ifx \bsnm \undefined \def \bsnm#1{#1}\fi
\ifx \bsuffix \undefined \def \bsuffix#1{#1}\fi
\ifx \bparticle \undefined \def \bparticle#1{#1}\fi
\ifx \barticle \undefined \def \barticle#1{#1}\fi
\bibcommenthead
\ifx \bconfdate \undefined \def \bconfdate #1{#1}\fi
\ifx \botherref \undefined \def \botherref #1{#1}\fi
\ifx \url \undefined \def \url#1{\textsf{#1}}\fi
\ifx \bchapter \undefined \def \bchapter#1{#1}\fi
\ifx \bbook \undefined \def \bbook#1{#1}\fi
\ifx \bcomment \undefined \def \bcomment#1{#1}\fi
\ifx \oauthor \undefined \def \oauthor#1{#1}\fi
\ifx \citeauthoryear \undefined \def \citeauthoryear#1{#1}\fi
\ifx \endbibitem  \undefined \def \endbibitem {}\fi
\ifx \bconflocation  \undefined \def \bconflocation#1{#1}\fi
\ifx \arxivurl  \undefined \def \arxivurl#1{\textsf{#1}}\fi
\csname PreBibitemsHook\endcsname

\bibitem[\protect\citeauthoryear{Wanten et~al.}{2023}]{wanten_plasma-based_2023}
\begin{barticle}
\bauthor{\bsnm{Wanten}, \binits{B.}},
\bauthor{\bsnm{Vertongen}, \binits{R.}},
\bauthor{\bsnm{De~Meyer}, \binits{R.}},
\bauthor{\bsnm{Bogaerts}, \binits{A.}}:
\batitle{Plasma-based {CO2} conversion: {How} to correctly analyze the performance?}
\bjtitle{Journal of Energy Chemistry}
\bvolume{86},
\bfpage{180}--\blpage{196}
(\byear{2023})
\doiurl{10.1016/j.jechem.2023.07.005}
\end{barticle}
\endbibitem

\bibitem[\protect\citeauthoryear{Sun et~al.}{2017}]{sun_co2_2017}
\begin{barticle}
\bauthor{\bsnm{Sun}, \binits{S.R.}},
\bauthor{\bsnm{Wang}, \binits{H.X.}},
\bauthor{\bsnm{Mei}, \binits{D.H.}},
\bauthor{\bsnm{Tu}, \binits{X.}},
\bauthor{\bsnm{Bogaerts}, \binits{A.}}:
\batitle{{CO2} conversion in a gliding arc plasma: {Performance} improvement based on chemical reaction modeling}.
\bjtitle{Journal of CO2 Utilization}
\bvolume{17},
\bfpage{220}--\blpage{234}
(\byear{2017})
\doiurl{10.1016/j.jcou.2016.12.009}
\end{barticle}
\endbibitem

\bibitem[\protect\citeauthoryear{Ramakers et~al.}{2017}]{ramakers_gliding_2017}
\begin{barticle}
\bauthor{\bsnm{Ramakers}, \binits{M.}},
\bauthor{\bsnm{Trenchev}, \binits{G.}},
\bauthor{\bsnm{Heijkers}, \binits{S.}},
\bauthor{\bsnm{Wang}, \binits{W.}},
\bauthor{\bsnm{Bogaerts}, \binits{A.}}:
\batitle{Gliding {Arc} {Plasmatron}: {Providing} an {Alternative} {Method} for {Carbon} {Dioxide} {Conversion}}.
\bjtitle{ChemSusChem}
\bvolume{10}(\bissue{12}),
\bfpage{2642}--\blpage{2652}
(\byear{2017})
\doiurl{10.1002/cssc.201700589}
\end{barticle}
\endbibitem

\bibitem[\protect\citeauthoryear{Moss et~al.}{2017}]{moss_investigation_2017}
\begin{barticle}
\bauthor{\bsnm{Moss}, \binits{M.S.}},
\bauthor{\bsnm{Yanallah}, \binits{K.}},
\bauthor{\bsnm{Allen}, \binits{R.W.K.}},
\bauthor{\bsnm{Pontiga}, \binits{F.}}:
\batitle{An investigation of {CO}$_{\textrm{2}}$ splitting using nanosecond pulsed corona discharge: effect of argon addition on {CO}$_{\textrm{2}}$ conversion and energy efficiency}.
\bjtitle{Plasma Sources Science and Technology}
\bvolume{26}(\bissue{3}),
\bfpage{035009}
(\byear{2017})
\doiurl{10.1088/1361-6595/aa5b1d}
\end{barticle}
\endbibitem

\bibitem[\protect\citeauthoryear{Zhang et~al.}{2022}]{zhang_boosting_2022}
\begin{barticle}
\bauthor{\bsnm{Zhang}, \binits{H.}},
\bauthor{\bsnm{Tan}, \binits{Q.}},
\bauthor{\bsnm{Huang}, \binits{Q.}},
\bauthor{\bsnm{Wang}, \binits{K.}},
\bauthor{\bsnm{Tu}, \binits{X.}},
\bauthor{\bsnm{Zhao}, \binits{X.}},
\bauthor{\bsnm{Wu}, \binits{C.}},
\bauthor{\bsnm{Yan}, \binits{J.}},
\bauthor{\bsnm{Li}, \binits{X.}}:
\batitle{Boosting the {Conversion} of {CO2} with {Biochar} to {Clean} {CO} in an {Atmospheric} {Plasmatron}: {A} {Synergy} of {Plasma} {Chemistry} and {Thermochemistry}}.
\bjtitle{ACS Sustainable Chemistry \& Engineering}
\bvolume{10}(\bissue{23}),
\bfpage{7712}--\blpage{7725}
(\byear{2022})
\doiurl{10.1021/acssuschemeng.2c01778}
\end{barticle}
\endbibitem

\bibitem[\protect\citeauthoryear{Li et~al.}{2019}]{li_plasma-assisted_2019}
\begin{barticle}
\bauthor{\bsnm{Li}, \binits{L.}},
\bauthor{\bsnm{Zhang}, \binits{H.}},
\bauthor{\bsnm{Li}, \binits{X.}},
\bauthor{\bsnm{Kong}, \binits{X.}},
\bauthor{\bsnm{Xu}, \binits{R.}},
\bauthor{\bsnm{Tay}, \binits{K.}},
\bauthor{\bsnm{Tu}, \binits{X.}}:
\batitle{Plasma-assisted {CO2} conversion in a gliding arc discharge: {Improving} performance by optimizing the reactor design}.
\bjtitle{Journal of CO2 Utilization}
\bvolume{29},
\bfpage{296}--\blpage{303}
(\byear{2019})
\doiurl{10.1016/j.jcou.2018.12.019}
\end{barticle}
\endbibitem

\bibitem[\protect\citeauthoryear{Fenghour et~al.}{1998}]{fenghour_viscosity_1998}
\begin{botherref}
\oauthor{\bsnm{Fenghour}, \binits{A.}},
\oauthor{\bsnm{Wakeham}, \binits{W.A.}},
\oauthor{\bsnm{Vesovic}, \binits{V.}}:
The {Viscosity} of {Carbon} {Dioxide}.
J. Phys. Chem. Ref. Data
\textbf{27}(1)
(1998)
\end{botherref}
\endbibitem

\bibitem[\protect\citeauthoryear{Anwar and Carroll}{2016}]{anwar_carbon_2016}
\begin{bbook}
\bauthor{\bsnm{Anwar}, \binits{S.}},
\bauthor{\bsnm{Carroll}, \binits{J.}}:
\bbtitle{Carbon {Dioxide} {Thermodynamic} {Properties} {Handbook}: {Covering} {Temperatures} from -20° to 250°{C} and {Pressures} up to 1000 {Bar}: {Second} {Edition}},
(\byear{2016}).
\doiurl{10.1002/9781119083948}
\end{bbook}
\endbibitem

\bibitem[\protect\citeauthoryear{Pope}{2000}]{pope_turbulent_2000}
\begin{bbook}
\bauthor{\bsnm{Pope}, \binits{S.B.}}:
\bbtitle{Turbulent Flows}.
\bpublisher{Cambridge University Press},
\blocation{Cambridge}
(\byear{2000})
\end{bbook}
\endbibitem

\bibitem[\protect\citeauthoryear{Abdel-Rahman}{2010}]{abdel-rahman_review_2010}
\begin{botherref}
\oauthor{\bsnm{Abdel-Rahman}, \binits{A.}}:
A {Review} of {Effects} of {Initial} and {Boundary} {Conditions} on {Turbulent} {Jets}.
WSEAS Transactions on Fluid Mechanics
\textbf{5}
(2010)
\end{botherref}
\endbibitem

\bibitem[\protect\citeauthoryear{Fellouah et~al.}{2009}]{fellouah_reynolds_2009}
\begin{barticle}
\bauthor{\bsnm{Fellouah}, \binits{H.}},
\bauthor{\bsnm{Ball}, \binits{C.G.}},
\bauthor{\bsnm{Pollard}, \binits{A.}}:
\batitle{Reynolds number effects within the development region of a turbulent round free jet}.
\bjtitle{International Journal of Heat and Mass Transfer}
\bvolume{52}(\bissue{17}),
\bfpage{3943}--\blpage{3954}
(\byear{2009})
\doiurl{10.1016/j.ijheatmasstransfer.2009.03.029}
\end{barticle}
\endbibitem

\bibitem[\protect\citeauthoryear{Xu and Antonia}{2002}]{xu_effect_2002}
\begin{barticle}
\bauthor{\bsnm{Xu}, \binits{G.}},
\bauthor{\bsnm{Antonia}, \binits{R.A.}}:
\batitle{Effect of different initial conditions on a turbulent round free jet}.
\bjtitle{Experiments in Fluids}
\bvolume{33},
\bfpage{677}--\blpage{683}
(\byear{2002})
\doiurl{10.1007/s00348-002-0523-7}
\end{barticle}
\endbibitem

\bibitem[\protect\citeauthoryear{Kwon and Seo}{2005}]{kwon_reynolds_2005}
\begin{barticle}
\bauthor{\bsnm{Kwon}, \binits{S.J.}},
\bauthor{\bsnm{Seo}, \binits{I.W.}}:
\batitle{Reynolds number effects on the behavior of a non-buoyant round jet}.
\bjtitle{Experiments in Fluids}
\bvolume{38},
\bfpage{801}--\blpage{812}
(\byear{2005})
\doiurl{10.1007/s00348-005-0976-6}
\end{barticle}
\endbibitem

\bibitem[\protect\citeauthoryear{Liu et~al.}{2019}]{liu_insight_2019}
\begin{barticle}
\bauthor{\bsnm{Liu}, \binits{J.-B.}},
\bauthor{\bsnm{Li}, \binits{X.-S.}},
\bauthor{\bsnm{Liu}, \binits{J.-L.}},
\bauthor{\bsnm{Zhu}, \binits{A.-M.}}:
\batitle{Insight into gliding arc ({GA}) plasma reduction of {CO2} with {H2}: {GA} characteristics and reaction mechanism}.
\bjtitle{Journal of Physics D: Applied Physics}
\bvolume{52}(\bissue{28}),
\bfpage{284001}
(\byear{2019})
\doiurl{10.1088/1361-6463/ab1bb1}
\end{barticle}
\endbibitem

\bibitem[\protect\citeauthoryear{Nagassou et~al.}{2019}]{nagassou_solargliding_2019}
\begin{barticle}
\bauthor{\bsnm{Nagassou}, \binits{D.}},
\bauthor{\bsnm{Mohsenian}, \binits{S.}},
\bauthor{\bsnm{Bhatta}, \binits{S.}},
\bauthor{\bsnm{Elahi}, \binits{R.}},
\bauthor{\bsnm{Trelles}, \binits{J.P.}}:
\batitle{Solar–gliding arc plasma reactor for carbon dioxide decomposition: {Design} and characterization}.
\bjtitle{Solar Energy}
\bvolume{180},
\bfpage{678}--\blpage{689}
(\byear{2019})
\doiurl{10.1016/j.solener.2019.01.070}
\end{barticle}
\endbibitem

\bibitem[\protect\citeauthoryear{Priyadarshini et~al.}{2024}]{priyadarshini_perception_2024}
\begin{barticle}
\bauthor{\bsnm{Priyadarshini}, \binits{M.S.}},
\bauthor{\bsnm{Bajaj}, \binits{M.}},
\bauthor{\bsnm{Prokop}, \binits{L.}},
\bauthor{\bsnm{Berhanu}, \binits{M.}}:
\batitle{Perception of power quality disturbances using {Fourier}, {Short}-{Time} {Fourier}, continuous and discrete wavelet transforms}.
\bjtitle{Scientific Reports}
\bvolume{14},
\bfpage{3443}
(\byear{2024})
\doiurl{10.1038/s41598-024-53792-9}
\end{barticle}
\endbibitem

\bibitem[\protect\citeauthoryear{}{}]{noauthor_short-time_nodate}
\begin{botherref}
Short-{Time} {Fourier} {Transform} - an overview {\textbar} {ScienceDirect} {Topics}
\end{botherref}
\endbibitem

\end{thebibliography}
\end{document}


\newpage
\section{Supporting Information}
\subsection{Performance Metrics - Equations}
CO$_2$ conversion was calculated from mass spectrometer signal intensities normalized across the species of interest in the chemical reaction (R1).
\begin{equation}
\mathrm{CO_2 \rightleftharpoons  CO + \tfrac{1}{2}O_2}
\tag{R1}
\end{equation}
A normalized signal intensity (Z\textsubscript{CO$_2$}) was defined as follows:
\begin{equation}
Z_{\mathrm{CO_2}} =
\frac{S_{\mathrm{CO_2}}/\mathrm{RSF}_{CO_2}}
{\sum_{i=1}^{3} \left( S_i / \mathrm{RSF}_i \right)}
\label{eq:zco2}
\tag{E1}
\end{equation}
where 'i' denotes chemical species from \textbf{R1}, S$_i$ is the measured signal at mass-to-charge ratio (m/z)$_i$, and RSF$_{i}$ is the corresponding relative sensitivity factor calibrated for each species. Raw conversion (X$_{raw}$) is defined relative to the plasma-on and plasma-off measurements. 

\begin{equation}
X_{\rm raw} = \frac{Z_{\rm CO_2,\, off} - Z_{\rm CO_2,\,on}}{Z_{\rm CO_2,\,off}}
\tag{E2}
\end{equation}
Because the chemical reaction increases the number of gaseous moles, and mass spectrometry is a concentration-based measurement technique, raw conversion overestimates the true conversion. A dilution correction following the work conducted by Wanten et al. gives the true conversion (X).\cite{wanten_plasma-based_2023}
\begin{equation}
X\,(\%) = \frac{2\, \times X_{\rm raw}}{3 - X_{\rm raw}} \times 100
\tag{E3}
\end{equation}
Plasma reactors are generally characterized by either plug-power, the total power drawn by the supply, or plasma-power, the power delivered to the reactor. Plasma power is reported throughout this work to directly assess the power delivered towards driving the chemical reactivity under study as opposed to plug power.  GAD CO$_2$ splitting performance is conventionally reported using three metrics: Energy Efficiency (EE), Energy Cost (EC), and Production Rate (PR$_{\mathrm{CO}}$), with specific energy input (SEI) being used as a descriptor of energy investment per unit volume. Using injected flow rate $\dot{Q}^{\mathrm{in}}_{\mathrm{CO_2}}$ (L/min), standard enthalpy of reaction ($\Delta H^{\circ} = 283~\mathrm{kJ/mol}$), and molar volume ($V_{\mathrm{m}} = 24.06~\mathrm{L/mol}$ at 293 K and 1 atm). 
\begin{equation}
\mathrm{EE}(\%) =
\frac{ X(\%) \times \Delta H^{\circ} \times \dot{Q}^{\mathrm{in}}_{\mathrm{CO_2}}}
     { P_{\mathrm{plasma}}~\mathrm{(kW)} \times V_{\mathrm{m}} \times 60~\mathrm{(s/min)} }
\label{EE}
\tag{E4}
\end{equation}

\begin{equation}
\mathrm{SEI~(MJ/L)} = 
\frac{ P_{\mathrm{plasma}} \times 60~\mathrm{(s/min)} }
     { Q^{\mathrm{in}}_{\mathrm{CO_2}} \times 1000~\mathrm{(kJ/MJ)} }
     \tag{E5}
\end{equation}

\begin{equation}
\mathrm{EC~(eV/mol.)} = 
\frac{ \mathrm{SEI} \times V_{\mathrm{m}} \times (6.242\times 10^{24})~\mathrm{(eV/MJ)} }
     { \left(\frac{X}{100}\right) \times N_{\mathrm{A}}~\mathrm{(mol./mol)} }
     \tag{E6}
\end{equation}

\begin{equation}
\mathrm{PR}_{\mathrm{CO}}~(\mathrm{g/h}) =
\frac{ \dot{Q}^{\mathrm{in}}_{\mathrm{CO_2}} \times X(\%) \times 28~\mathrm{(g/mol)} \times 60~\mathrm{(min/h)} }
     { V_{\mathrm{m}} \times 100~(\%) }
     \tag{E7}
\end{equation}

The choice of molar volume must be carefully considered to align with operating conditions as variations in temperature and pressure affect the calculated metrics.\cite{wanten_plasma-based_2023}

To contextualize this work within the broader gliding arc literature, \textbf{Table 1} compares operating scale, conversion, and energy cost for this reactor with other gliding arc and gliding arc plasmatron studies of CO$_2$ splitting. This places the present work in a comparable position relative to other studies, indicating that the system, while designed for diagnostic purposes, retains performance characteristics representative of GAD reactors more broadly.

\begin{table*}[h]
    \centering
    \caption{Typical Performance Metrics of Gliding Arc Discharges and  Plasmatrons}
    \label{tab:placeholder_label}
    \begin{tabular}{lcccccc}
        \toprule
        Reactor & $Q^{\mathrm{in}}_{\mathrm{CO_2}}$ (L/min) & X (\%) & \multicolumn{1}{c}{EC (eV/mol.)} & Ref. \\
        \textbf{GAD} & \textbf{1.5} & \textbf{3.2} & \textbf{13.1} & \textbf{This Work} \\
        GAD   & 6.5 & 9.6 & 9.0 & \cite{sun_co2_2017} \\
        GAP  & 10  & 8.6 & 9.8  &\ \cite{ramakers_gliding_2017} \\
        GAD   & 10  & 6   & 10.1 & \cite{moss_investigation_2017} \\
        GAP  & 10  & 7   & 10.3  & \cite{zhang_boosting_2022} \\
        GAD   & 3   & 11.1 & 14.2  & \cite{li_plasma-assisted_2019} \\
    \end{tabular}
\end{table*}

\subsection{Jet Expansion and Reynolds Number}
\subsubsection{Reynolds Number}
The Reynolds number (Re), represents the ratio of inertial to viscous forces in the flow and was calculated using the gas density and viscosity at the inlet nozzle, before plasma exposure. This distinction is important, as gas properties are functions of temperature and GADs exhibit large spatiotemporal thermal gradients across the arc. As such, referencing the Re before the plasma is essential for meaningful analysis.
\begin{equation}
\mathrm{Re} =
\frac{4 \times \rho \times \dot{Q}^{\mathrm{in}}}
{\pi \times D_{\mathrm{nozzle}} \times \mu}
\tag{E8}
\end{equation}
For pure CO$_2$ at 293K and 1 atm, $\rho$ = 1.795 kg/m$^3$, and $\mu$ = 1.47 x 10$^{-5}$ Pa$\cdot$s.\cite{fenghour_viscosity_1998,anwar_carbon_2016}

\subsubsection{Jet Envelope Calculation}
As gas exits a conflat nozzle, the jet diameter (D$_{jet}$), expands as a function of distance from the nozzle (z) according to classical fluid dynamic equations.\cite{pope_turbulent_2000}

\begin{equation}
D_{\mathrm{jet}}(z) =
\begin{cases}
D_{\mathrm{nozzle}}, & 0 < z < Z_0 \\[2mm]
2 \times S \, \times(z - Z_0) + D_{\mathrm{nozzle}}, & z \ge Z_0
\end{cases}
\tag{E9}
\end{equation}

Here $Z_0$ represents the distance from the nozzle that the gas begins to spread in the near field, and S is the spreading rate.~\cite{abdel-rahman_review_2010} Empirical factors of $Z_0$ = 2.5 × D$_{nozzle}$ and $S$ = 0.086 were taken from published works by Fellouah et al. and Xu et al., respectively.~\cite{fellouah_reynolds_2009,xu_effect_2002} Kwon et al. demonstrated stability of these equations and empirical parameters at Reynolds numbers above 2000, supporting the use of these relations across the flow conditions tested in this study.\cite{kwon_reynolds_2005}

\subsubsection{Jet-to-Plasma Overlap Calculation}
Jet-to-plasma (JtP) overlap is defined as the fraction of the inlet jet cross-section that intersects a plasma arc formed at the minimum electrode gap. The arc is treated as a thin plane, of height D$_{arc}$, passing through the center of the jet cross-section. The arc diameter for a pure CO$_2$ plasma arc was reported by Liu et al. to be $\approx$ 1 mm. \cite{liu_insight_2019} The shared area between the plane and circular jet cross-section is given by:

\begin{equation}
A_{\text{shared}} =
\begin{cases}
\dfrac{D_{\text{jet}}^{2}}{2}\times\sin^{-1}\!\left(\dfrac{D_{\text{arc}}}{D_{\text{jet}}}\right)
+
\dfrac{D_{\text{arc}}}{2}\times\sqrt{D_{\text{jet}}^{2}-D_{\text{arc}}^{2}},
& D_{\text{arc}} < D_{\text{jet}} \\[10pt]

\dfrac{\pi}{4}D_{\text{jet}}^{2},
& D_{\text{arc}} \ge D_{\text{jet}}
\end{cases}
\tag{E10}
\end{equation}
JtP overlap is then expressed as the percentage of shared area relative to the jet cross-section.
\begin{equation}
\text{JtP Overlap}(\%) =
\frac{A_{\text{shared}}}{\dfrac{\pi}{4}D_{\text{jet}}^{2}}
\times 100
\tag{E11}
\end{equation}

\clearpage

\begin{wrapfigure}[25]{l}{0.5\textwidth}
    \centering
    \includegraphics[width=0.45\textwidth]{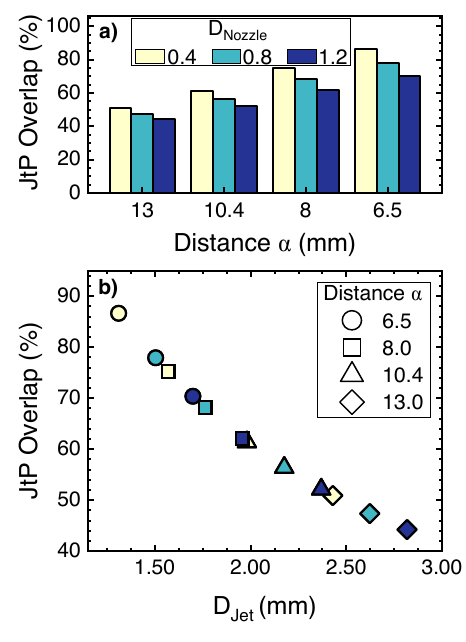}
    \caption{
Relationships between geometric setups of the GAD and JtP overlap. (a) JtP as a function of electrode-to-nozzle distance ($\alpha$) and nozzle sizes. (b) JtP overlap as a function of jet diameter as extracted from E9-E11.
    }
    \label{S1}
\end{wrapfigure}

JtP overlap represents an upper bound on the percentage of feed gas that passes through a plasma arc formed during a Mode A initiation event. Because gliding arcs are dynamic in morphology, travel path, and discharge time this geometric quantity does not capture true reactant-plasma exposure and should be interpreted only as a relative measure of potential reactant exposure. \textbf{Figure 1b} shows the relationship between electrode-to-nozzle distance, nozzle diameter, D$_{jet}$ and JtP overlap. The minimum electrode gap of 3.2 mm used in this study is substantially larger than the largest D$_{jet}$ tested, supporting that the influence of the electrodes on the flow field can be considered negligible.

\subsection{Additional Routes for Improved Performance}
The minimum inter-electrode gap of 3.2 mm used in this study was selected to keep the electrodes outside the inlet jet envelope and avoid potential wall effects affecting the flow environment. However, this large gap reduces the electron density of the arcs and thus reduces achievable conversion and energy efficiency, as described by Sun et al. and others.\cite{sun_co2_2017, li_plasma-assisted_2019} These studies demonstrated that decreasing the inter-electrode gap size can improve energy efficiency by nearly 10\%, showing there is substantial potential for GAD reactors to be driven using neon sign transformers and be directly competitive with research-grade plasma power supplies through improved reactor design.

\begin{wrapfigure}[19]{r}{0.5\textwidth}
    \centering
    \includegraphics[width=0.45\textwidth]{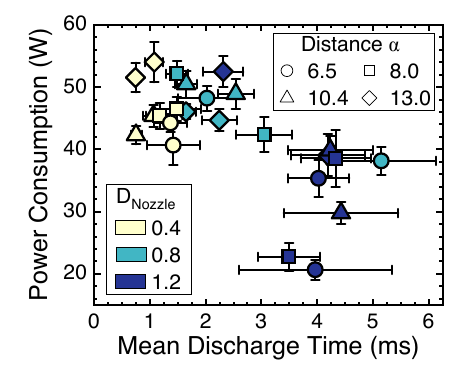}
    \caption{
Relationship between power consumption and mean discharge time across all reactor nozzle and electrode-to-nozzle distance combinations. Shorter mean discharge times correlate with higher power consumption, though with substantial spread.
    }
\end{wrapfigure}

While power consumption is often reported to describe reactor operation, this collective descriptor of the waveform does not provide direct insight into the underlying arc dynamics characterizing the reactor. \textbf{Figure 2} shows the relationship between power consumption and mean discharge time, where shorter times correlate with higher powers, but with substantial spread. This suggests that both power consumption and mean discharge time must be considered jointly to fully characterize GADs, with mean discharge time describing arc formation frequency and operating stability while power consumption describes propagation and total energy investment, which vary as a function of arc dynamics.

\clearpage
\subsection{High-Speed Image Processing}

Quantities in this work drawn from high-speed video, including the discharge-mode
classification, the optical mean discharge time, the mode-transition
probabilities, the maximum arc length per cycle, and the arc velocities was
obtained with the following image processing and analysis pipeline. Videos were acquired on a Phantom v2512 high-speed
camera at 200{,}000~fps (\SI{5}{\micro\second} frame interval) in a $256\times256$~px viewing window over sampled values of
nozzle diameter and CO$_2$ flow rate reported in the main text.

\subsubsection{Acquisition and spatial calibration}
The recordings contain no in-frame length reference, so the spatial scale was
fixed by the minimum inter-electrode
gap of 3.2~mm. For
each recording, the pixel-to-millimeter scale was derived per video from the
hand-segmented electrode tips. Each frame was captured as a raw .cine file, and later converted to .tif for analysis.

\subsubsection{Arc detection and per-frame measurement}
Arcs were isolated
with a fixed intensity threshold unique to each recording. The thresholded mask was morphologically closed, and only the single largest connected
component was retained. Area gates rejected saturated blobs, and
highly saturated electrode gap flashes were rejected. Each arc that made it to a retained mask was thinned into a one-pixel skeleton with the
Zhang--Suen thinning algorithm as implemented in
scikit-image, allowing for the arc length (total skeleton
length converted with px/mm) and the advancing-front
tip (the extreme mask pixel in the propagation direction) to be measured. A frame contributed to the arc statistics only if the arc was
detected, completely bridged the electrodes, and was not flagged as a
flash or blob by the flash and width gate filters. Within each discharge event, the per-frame arc length was tracked and its peak recorded
as the greatest path length the arc reached
before re-striking, called the maximum arc length.

\subsubsection{Optical mode classification}
Each initiation was assigned to either mode A or mode B (the only discharge modes discernible by optical data alone). The boundary to demarcate mode separation was assigned on a per-recording basis and was placed at the dead zone of the re-strike initiation depth distribution instead of an arbitrary distance from the minimum gap position. For each recording, the distribution of depths at which the leading edge of the arc re-initiates at or below the minimum gap is bimodal. One sharp peak lies near the minimum gap, and a second peak lies further down the electrode gap, separated by a sparse re-strike valley. The boundary depth is taken to be that interval of minimum initiation. This depth is found by binning a histogram of re-strike depth and taking the minimum of the envelope in the space between the two peaks. A new arc appearing above this boundary was labeled Mode~A and below Mode~B. Across the recordings, this boundary line varies between 3.8 and 5.4 mm below the minimum gap, with an average value of 4.4 mm and no clear trend in boundary depth between Re. An extended interval of $>$500 frames where no arc was detected was labeled Mode~C, though Mode C was only realized for the Re = 3239 recording.

\subsubsection{Optical discharge statistics}
New arc formation events were identified optically from frame-to-frame changes
in the arc's pixel position, measured by the leading arc tip position. The mean interval
between successive detected initiations gives the optical mean discharge time, which is compared against the electrical value (see \textbf{Figure 4}). This
detection responds to a change in pixel position and is limited to initiations that displace the
arc beyond the camera's spatial resolution. This accounts for
the deviation from the electrical value at the lowest Re value. The optical mode B prevalence reported against Re elsewhere in \textbf{Figure 6a} is the fraction of
initiation events classified as Mode~B for each condition. The conditional probability matrix, in \textbf{Figure 6b}, was built from the ordered
sequence of classified events by counting transitions between
successive initiations and normalizing each originating-mode row to unity, so that element $(i,j)$ is the probability that a re-strike initiation in mode i is followed by one in mode j. Reset intervals enter the sequence as Mode~C events rather than being excluded. Counts were pooled across conditions before normalization, so the matrix is weighted by sample size. The unstable Re shows some statistical differences in conditional probability in favoring mode transition more heavily than its counterparts, which show more independent mode selection.

\subsubsection{Arc velocities}
Two velocities with distinct physical meaning were computed from the tracked tip. The advance velocity is event-averaged: the tip position over propagating frames within an event was fit with a linear regression, and the fitted slope was converted to a physical velocity by \textbf{E12}.

\begin{equation}
v~[\mathrm{m/s}] = \mathrm{slope}\times\frac{\mathrm{fps}}{(\mathrm{px/mm})\times 1000},
\tag{E12}
\end{equation}

\begin{wrapfigure}[23]{l}{0.5\textwidth}
\centering
\includegraphics[width=0.45\textwidth, keepaspectratio]{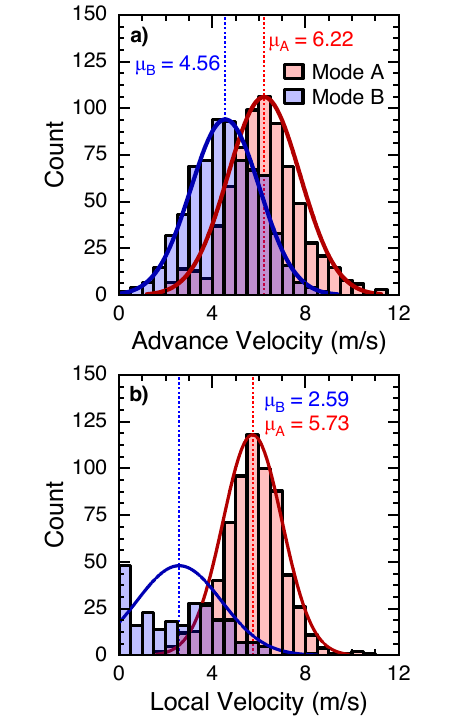}
\caption{
Velocity separation of Mode A and B initiation events using both advance and local velocities at Re 3239.
}
\end{wrapfigure}

The slope represents the fitted tip displacement in pixels per frame, and $v$ represents the mean rate at which the front sweeps along the electrodes over an entire initiation-extinction event. The local tip velocity is a time-resolved descriptor. It is the slope of a centered sliding-window fit of tip position over a $\pm 5$ frame window (each point plus its five neighbors on either side), converted to velocity with the same relation. The local tip velocity was used to build spatially resolved maps of arc velocity. It physically represents the front kinematics rather than event-averaged rate. Local tip velocity was only used to create velocity heat maps, due to its relation to spatial behaviors of the entire arc body, while advance velocity was used analyze collective trends of the overarching arc behavior.

\clearpage
\subsection{Quantification of Arc Dynamics and Behaviors}
\subsubsection{Electrical and Optical Mean Discharge Times}
\begin{wrapfigure}[21]{l}{0.5\textwidth}
\centering
\includegraphics[width=0.45\textwidth,keepaspectratio]{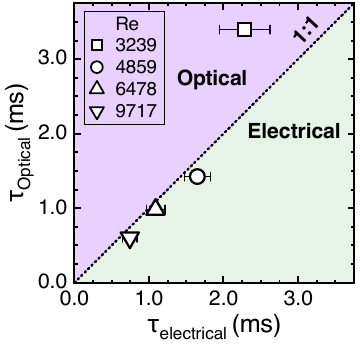}
\caption{
Comparisons of mean discharge time ($\tau$) from electrical and optical methods across a range of Re demonstrating strong overlap when Re is above 4000.
}
\end{wrapfigure}

For comparison between optical and electrical event results, mean discharge times have been computed for both approaches in \textbf{Figure 4}. Optical mean discharge times are computed from high-speed images by tracking shifts in arc's leading pixel position to identify new arc-formation events. Electrical event detection, in contrast, only requires sufficiently high temporal resolution of the electrical waveform to separate plasma events rather than visual confirmation. Electrically, these events have been further sub-categorized as Modes A-D as previously described.

Of particular importance to this comparison is Mode B, in which a new arc forms away from the minimum electrode position and there is a disruption in the current waveform at the time of arc formation, and Mode D, where visually it appears that a new arc has formed, but there is no statistically detectable change in the current. Because the electrical mean discharge time explicitly tracks Mode A and B events and separates out Mode D contributions, it yields longer average electrical discharge times than the optical approach, which cannot distinguish these events. 

The optical and electrical mean discharge times agree closely for Re above 4000, while the smallest Re evaluated shows substantial deviation from a 1:1 relationship. This deviation arises from a sensitivity limitation of the optical diagnostics: because new arc formation relies on changes in pixel intensity by location to classify new arc formation, they are limited by the spatial resolution provided by the camera,  particularly when new arc formations occur with minimal changes in pixel position. 

Fluctuations in pixel intensity due to AC-nature of the supply also led to difficulty distinguishing between frames without arcs and frames with faint arcs that did not make it past image processing filters (see \textbf{S1.4} for details). 

Such events remain detectable electrically, as any shortening in the arc volume creates a voltage drop and can have an associated change in the current signal, which is the only basis required to classify an event. Together, these results indicate that the electrical signature tracking developed in this work provides a more robust approach for monitoring plasma dynamics than optical tracking alone.

\subsubsection{Power-Law Scaling and Conditional Probabilities}
The data demonstrates a strong decaying power law relationship between Re and electrical mean discharge time as shown in \textbf{Figure 5}. This relationship was derived from the electrical event detection compiled across 30 individual waveforms at each condition, using the aggregated mean discharge time from each electrical waveform. 

Importantly, reactor orientation has been reported to affect arc speed and must therefore be considered when reporting these characteristic scaling relationships.\cite{nagassou_solargliding_2019} This work was conducted such that the reactor was oriented vertically with arc propagation in alignment with gravity, thereby opposing buoyancy forces.

\clearpage
\begin{wrapfigure}[20]{l}{0.45\textwidth}
\centering
\includegraphics[width=0.45\textwidth, keepaspectratio]{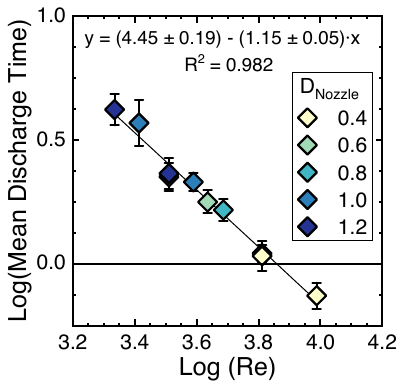}
\caption{
Power law relationship of Re and mean discharge time. Linear fit to log-log plot shows a strong fit to a power law model.
}
\end{wrapfigure}

Another notable question in the field of gliding arc discharges is how the prevalence of arc formation events that occur at the minimum electrode position versus elsewhere, denoted Mode A and B events in this work respectively, changes in relation to flow. The ratio of these events has been studied using both electrical and optical diagnostics from separate experiments, showing no strong relationship to Re from either dataset as presented in \textbf{Figure 6a}. This result indicates that the ratio of these events is largely shaped by the power supply or the reactor geometry, rather than the flow conditions in this study. 

Additionally, an investigation of conditional probabilities was conducted to determine whether coupling behaviors could be observed between different mode events. The results shown in \textbf{Figure 6b} indicate a statistical preference for swapping initiation mode types rather than cascades of a single modal type. In contrast, Mode C events, which arise from the secondary-circuit ground fault protection system in the NST creating intervals of no plasma, always result in a Mode A initiation event. This is consistent with Mode A representing the shortest ionization pathway when there are no pre-ionized molecules from a former arc accessible to allow for a Mode B event. Notably, this also skews the statistics presented in \textbf{Figure 6a} by biasing the Mode A:B ratio.

\begin{figure}[h]
\centering
\includegraphics[width=0.9\textwidth, keepaspectratio]{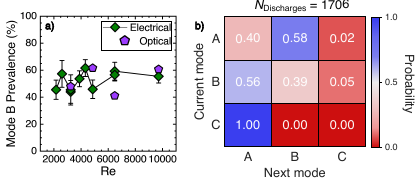}
\caption{
Prevalence of Mode B initiation events is largely invariant of Re and initiation events demonstrate greater likelihood of swapping initiation modal type. (a) Mode B prevalence as a function of Re from separate optical and electrical diagnostic experiments. (b) Conditional probability matrix of mode transitions of the plasma arc collected across all optically studied Re.
}
\end{figure}

Importantly, the analysis shown in \textbf{Figure 6b} was obtained from optical measurements rather than purely electrical measurements, since the oscilloscope's time-span limitations cap the analysis window to a few hundred milliseconds per recording, compared to the tens of seconds achievable with the high speed camera. As such, the optical analysis allows for a larger continuous sample size when screening conditional probabilities using the previously described criteria for modal labeling and sequence tracking. As Mode D requires joint optical and electrical analysis to identify, and is not of particular interest for this study, the associated probabilities of this event were not tabulated.

\clearpage
\subsection{Short-Time Fourier Transform on Voltage}
\begin{wrapfigure}[20]{r}{0.3\textwidth}
    \centering
    \includegraphics[width=\linewidth]{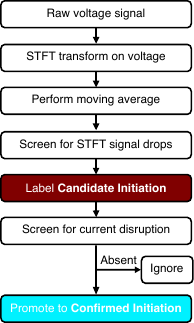}
    \caption{
    Simplified depiction of the computational tools workflow for electrical event detection.
    }
    \label{S1}
\end{wrapfigure}

A short-time Fourier transform (STFT) was applied to the raw voltage signal to develop code for finding voltage drops according to \textbf{E13} and \textbf{E14}.

\begin{equation}
\mathrm{V'_{STFT}}(\tau,\omega) 
= \int_{-\infty}^{\infty} V(t)\times W(t-\tau)\times e^{-i\omega t} dt
\tag{E13}
\end{equation}

Where $V(t)$ is the voltage signal, $W(t-\tau)$ is a window function centered at time index $\tau$, $t$ is time, and $\omega$ is frequency.\cite{priyadarshini_perception_2024,noauthor_short-time_nodate} These equations enable a finite Fourier transform to be calculated at each sliding window position across the entire voltage waveform. To identify voltage drops, a frequency-independent signal intensity was calculated as follows:
\begin{equation}
\mathrm{V_{STFT}}(\tau) = \frac{1}{N_\omega} \times\sum_{k=1}^{N_\omega} [ \mathrm{V'_{STFT}}(\omega_k,\tau)] ^2
\tag{E14}
\end{equation}

This equation represents the average V$_{STFT}$ magnitude across all frequencies at a fixed time, wherein N$_\omega$, represents the number of discrete frequency values. This calculation removes frequency dependence from the V'$_{STFT}$ signal and produces a time-dependent signal that directly reflects changes in the V(t) waveform.  
\subsubsection{Electrical Event Thresholding}
Computational screening of voltage drops through V$_{STFT}$ was achieved using a 512-sample window with 75\% overlap between consecutive windows. The resulting intensity was tracked by grouping peaks into runs, with a run continuing as long as intensity remained within 70\% of the prior value or continued increasing. A run ended once intensity dropped below 30\% of the run's peak value and remained low for 3 consecutive points, flagging a new candidate event.
\begin{figure*}[h]
\centering
\includegraphics[width=0.9\linewidth, keepaspectratio]{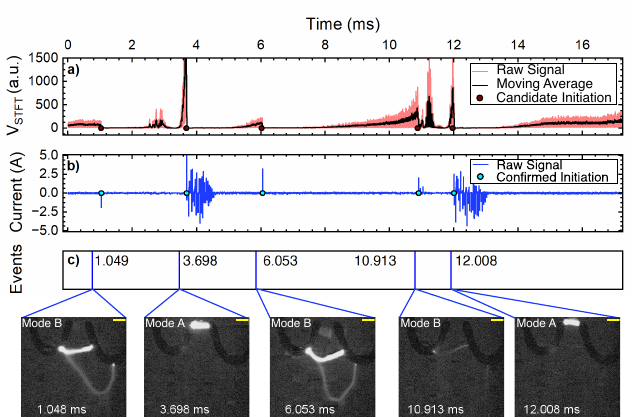}
\caption{
Electrical event detection aligns with high-speed imaging. (a) STFT of voltage identifies candidate initiation events spanning modes A–D. (b) Joint voltage and current analysis confirms mode A and B events. (c) Confirmed event timestamps with corresponding high-speed photography frames. Yellow scale bars represent 3.2 mm.
}
\end{figure*}
Each candidate was then checked against the current signal within a ±125 $\mu$s window to account for the phase delay between voltage and current measurements, and was promoted to a confirmed event only if a current spike of at least 175 mA was found nearby. This triggering current threshold was selected as it sits above one standard deviation of the propagating current. Mode determination for each confirmed event was based on the number of current disruptions above roughly 306 mA, 1.75 times the triggering current threshold, within a 275 $\mu$s forward-looking window. Events with 4 or fewer disruptions were classified as Mode B, and events with more than 4 were classified as Mode A. The full implementation, including all thresholding parameters, is provided in the associated code repository.

\begin{figure*}[h]
\centering
\includegraphics[width=0.9\linewidth, keepaspectratio]{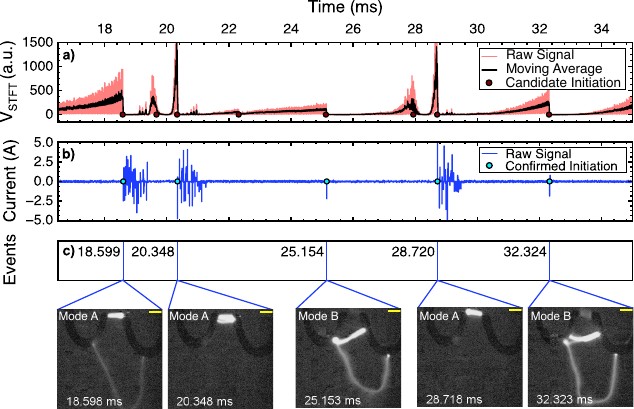}
\caption{
Continuation of \textbf{Figure 8} for electrical event detection. (a) STFT of voltage identifies candidate initiation events spanning modes A–D. (b) Joint voltage and current analysis confirms mode A and B events. (c) Confirmed event timestamps with corresponding high-speed photography frames. Yellow scale bars represent 3.2 mm.
}
\end{figure*}
\begin{figure*}[t]
\centering
\includegraphics[width=0.9\linewidth, keepaspectratio]{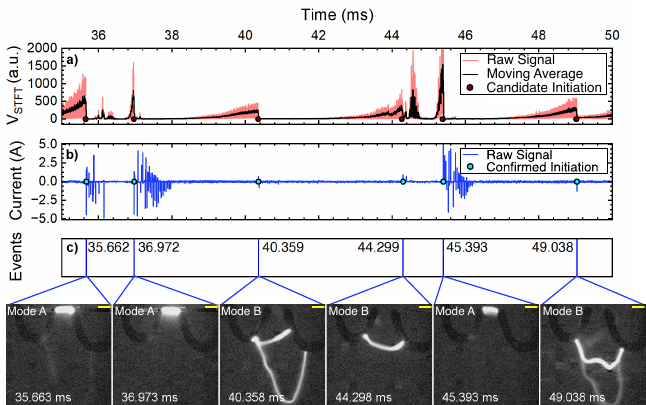}
\caption{
Continuation of \textbf{Figure 8} for electrical event detection. (a) STFT of voltage identifies candidate initiation events spanning modes A–D. (b) Joint voltage and current analysis confirms mode A and B events. (c) Confirmed event timestamps with corresponding high-speed photography frames. Yellow scale bars represent 3.2 mm.
}
\end{figure*}

\clearpage
\bibliography{MethodsTrimmed}